%% file: main.tex
\documentclass[a4paper,10pt]{article}
\usepackage{jheppubMOD} % for details on the use of the package, please
\usepackage{lipsum}
\usepackage{graphicx}  % needed for figures
\usepackage{multirow}
\usepackage[x11names,dvipsnames]{xcolor}

\usepackage{lmodern}
\usepackage{amsmath,amssymb}
\usepackage{mathtools}
\usepackage[numbers,sort&compress]{natbib}

\usepackage{tikz}
\usepackage{tikz-3dplot}
\usepackage{pgfplots,pgfplotstable}
\usepgfplotslibrary{groupplots}
\usetikzlibrary{decorations.markings}
\usetikzlibrary{shapes}
\usetikzlibrary{shapes.misc}
\usetikzlibrary{patterns}
\usepgfplotslibrary{fillbetween}
\pgfplotsset{compat=newest}
\usetikzlibrary{positioning}
\usetikzlibrary{calc}

\usepackage{lipsum}
\usepackage{stackrel}

\usepackage[compat=1.1.0]{tikz-feynman}

\usepackage{cleveref}
\usepackage{ifthen}

\usepackage[shortlabels]{enumitem}

\input{TikZsettings.tex}

\input{shorthands.tex}

\usepackage{lipsum}
\usepackage{graphicx}  % needed for figures
\usepackage{multirow}
\usepackage[x11names,dvipsnames]{xcolor}

\usepgfplotslibrary{external}
\usetikzlibrary{external}
\pgfrealjobname{LHCbTcc}

\graphicspath{{FiguresPDF/}}

\hypersetup{
    colorlinks,
    linkcolor={Blue},
    citecolor={Blue},
    urlcolor={Blue}
}

\usepackage{lmodern}
\usepackage{amsmath,amssymb}
\usepackage{mathtools}
\usepackage[numbers,sort&compress]{natbib}

\usepackage{stackrel}
\usepackage{bm}

\usepackage{cleveref}
\usepackage{ifthen}
\usepackage{xspace}
\usepackage[shortlabels]{enumitem}

\usepackage[left=2cm,right=2cm,top=3cm,bottom=3cm]{geometry}

\newcommand{\be}{\begin{equation}}
\newcommand{\ee}{\end{equation}}
\newcommand{\bea}{\begin{eqnarray}}
\newcommand{\eea}{\end{eqnarray}}
\newcommand{\bef}{\begin{figure}}
\newcommand{\eef}{\end{figure}}
\newcommand{\bce}{\begin{center}}
\newcommand{\ece}{\end{center}}

\newcommand{\ket}[1]{\left\lvert #1 \right\rangle}

\usepackage{bbm}

\newcommand{\DsC}{D^{\ast+}}
\newcommand{\DsN}{D^{\ast0}}
\newcommand{\DC}{D^+}
\newcommand{\DN}{D^0}
\newcommand{\cho}{\DsC\DN}
\newcommand{\cht}{\DsN\DC}
\newcommand{\CIsos}[3]{C_{D^{\ast{#1}} \to D^{#2} \pi^{#3}}}
\newcommand{\Nevconv}{\mathcal{\overline{N}}_\text{ev}}
\newcommand{\Nevnotconv}{\mathcal{N}_\text{ev}}

\newcommand{\mDsSt}{m_{D^\ast_{(t)}}^2}
\newcommand{\mDsSu}{m_{D^\ast_{(u)}}^2}

\newcommand{\Tccp}{T_{cc}^{+}}
\newcommand{\Tccpp}{T_{cc}^{++}}
\newcommand{\Tccz}{T_{cc}^{0}}
\newcommand{\TccSpp}{T_{cc}^{\ast++}}
\newcommand{\TccSp}{T_{cc}^{\ast+}}
\newcommand{\TccSz}{T_{cc}^{\ast0}}
\newcommand{\LHCb}{LHCb }

\newboolean{CompileDiagrams}
\setboolean{CompileDiagrams}{false}

\newboolean{CompileFigures}
\setboolean{CompileFigures}{false}

\begin{document}

\title{\boldmath $T_{cc}^{+}$ coupled channel analysis and predictions}

\newcommand{\ific}{Instituto de F\'isica Corpuscular (IFIC),
             Centro Mixto CSIC-Universidad de Valencia,
             Institutos de Investigaci\'on de Paterna,
             Aptd. 22085, E-46071 Valencia, Spain}

\author[a]{M.~Albaladejo}

\affiliation[a]{\ific}

\emailAdd{Miguel.Albaladejo@ific.uv.es}

\date{\today}

\abstract{A coupled channel analysis of the $D^{\ast+}D^0$ and $D^{\ast0}D^+$ system is performed to study the doubly charmed $T_{cc}^+$ state recently discovered by the LHCb collaboration. We use a simple model for the scattering amplitude that allows us to describe well the experimental spectrum, and obtain the $T_{cc}^+$ pole in the coupled channel $T$-matrix. We find that this bound state has a large molecular component. The isospin ($I=0$ or $I=1$) of the state cannot be inferred from the $D^0 D^0 \pi^+$ spectrum alone. Therefore, we use the same formalism to predict other $DD\pi$ spectra. In the case the $T_{cc}^+$ has $I=1$, we also predict the location of the other two members ($T_{cc}^{++}$ and $T_{cc}^0$) of the triplet. Finally, using Heavy-Quark Spin Symmetry, we predict the location of possible heavier $D^\ast D^\ast$ ($I=0$ or $I=1$) partners.}

\frenchspacing
%\toccontinuoustrue
\maketitle

\section{Introduction}

It is a fact that most of the discovered hadrons can be classified as $q\bar{q}$ (mesons) and $qqq$ (baryons) states within the constituent quark model \cite{Gell-Mann:1964ewy,Zweig:1964ruk,Zweig:1964jf,Godfrey:1985xj}. However, nothing prevents the existence of other color-singlet states with more complicated constituent quark and gluon structures. Besides being interesting in its own right, the finding of such states can give us precious insights into QCD, the fundamental underlying theory of strong interactions.

The \LHCb collaboration has reported a very prominent signal in the $\DN\DN\pi^+$ spectrum \cite{LHCb:2021vvq,LHCb:2021auc}, and has claimed the existence of a new state, named $\Tccp$. Given its decay channel, this state would have two charm-quarks. If confirmed in different experiments and/or by other collaborations, it would be the first meson with such features. The $\Xi_{cc}^{++}$ was the first and still only discovered baryon \cite{LHCb:2017iph} with doubly charm flavour. Due to its $cc$ quark content, the $\Tccp$ state would be a clearly exotic one, but the question of its nature remains open. A detailed discussion encompassing the $\Tccp$ state (with plenty of references on previous works) and the \textit{molecular} vs. \textit{compact tetraquark} assignment can be found in Sec.~IV.A of Ref.~\cite{Dong:2021bvy}. As a chain reaction, this announcement has stimulated numerous works \cite{Li:2021zbw,Meng:2021jnw,Dai:2021wxi,Feijoo:2021ppq,Agaev:2021vur,Wu:2021kbu,Ling:2021bir,Chen:2021vhg,Yan:2021wdl,Weng:2021hje,Huang:2021urd,Chen:2021kad,Xin:2021wcr,Fleming:2021wmk,Chen:2021tnn,Azizi:2021aib,Ren:2021dsi,Yang:2021zhe,Jin:2021cxj,Hu:2021gdg,Chen:2021cfl,He:2021smz}, with different outcomes depending on the analysis. Just as a sample, Ref.~\cite{Dai:2021wxi} analyses the raw \LHCb spectrum and finds the origin of the $\Tccp$ to be a virtual state, while in Ref.~\cite{Feijoo:2021ppq} subtraction constants are tuned so as to have a bound state with the mass determined by the \LHCb collaboration. Lattice QCD simulations do not provide definite conclusions about a possible $\Tccp$ either \cite{Junnarkar:2018twb,Ikeda:2013vwa,Cheung:2017tnt,Francis:2018jyb}, although the situation seems more clear in the bottom sector (see {\it e.g.} Ref.~\cite{Leskovec:2019ioa} and references therein).

Besides its quark content, the $\Tccp$ mass and width are also interesting. The \LHCb collaboration reports two different determinations,
\begin{subequations}\begin{align}
\Delta M_{\Tccp} = M_{\Tccp} - m_{\DsC} - m_{\DN} 
& = -273(61)\ \text{keV}~,
& \Gamma_{\Tccp} 
& =  410(165)\ \text{keV} & (\text{Ref.~\cite{LHCb:2021vvq}})~, \label{eq:LHCbBW}\\
& = -360(40) \ \text{keV}~,\,
& 
& =  \hphantom{0}48(2)\hphantom{00}  \ \text{keV} & (\text{Ref.~\cite{LHCb:2021auc}}) \label{eq:LHCbPole}~.
\end{align}\end{subequations}
The first determination stems from the use of a standard Breit-Wigner (BW) parameterisation, while the second one is obtained when a unitarised BW profile is used. Because of its closeness to the $\cho$ threshold, the hypothesis of $\Tccp$ being a molecular state is quite appealing. More importantly, being so close to threshold renders mandatory to take into account coupled channel dynamics in any realistic analysis. Also, because the $\Tccp$ binding energy is small and similar to the experimental resolution ($\sim 400\ \text{keV}$), and its width is even smaller, it is also very important to convolute the theoretical spectrum with the experimental resolution. We take into account in this work all these aspects. The importance of a careful study of the $\Tccp$ width has been highlighted, for instance, in Refs.~\cite{Yan:2021wdl,Meng:2021jnw,Fleming:2021wmk,Ling:2021bir}.

In this work we present a coupled channel $T$-matrix analysis of $D^\ast D$ scattering ($\cho$ and $\cht$), and incorporate it into a $DD\pi$ mechanism production. This formalism is presented in Sec.~\ref{sec:model}. In Sec.~\ref{sec:results} we show our results, starting from the fit of the model (Subsec.~\ref{subsec:fits}) to reproduce the experimental $\DN\DN\pi^+$ spectrum by the \LHCb collaboration \cite{LHCb:2021vvq,LHCb:2021auc}. We find in Subsec.~\ref{subsec:poles} that the $\Tccp$ signal originates from a $D^\ast D$ bound state, and in Subsec.~\ref{subsec:molecular} we advocate for its molecular nature. In Sec.~\ref{sec:otherpredictions}  we use the previously determined scattering matrix and mechanism production to make some predictions. In particular, we compute other $DD\pi$ spectra in Subsec.~\ref{subsec:PredictionsSpectra}, and in Subsec.~\ref{subsec:HQSS} we predict potential Heavy-Quark Spin Symmetry partners of the $\Tccp$ state. Conclusions are given in Sec.~\ref{sec:conclusions}.

\section{Model}\label{sec:model}

The \LHCb $\DN\DN\pi^+$ spectrum \cite{LHCb:2021vvq} consists essentially of the $\Tccp$ signal, very close to the $\cho$ threshold, and a background that seems to originate from the opening of that threshold, too. Therefore it is reasonable to assume that all of the $DD\pi$ spectrum comes from $D^\ast D$ pairs. Furthermore, because of the small range around the $D^\ast D$ thresholds that the data span, it is also reasonable to consider that these $D^\ast D$ pairs are produced in an $S$-wave, thus with quantum numbers $J^P=1^+$, and that higher waves are supressed. Therefore, we treat the $DD\pi$ spectrum as a decay of an axial source $S$ with invariant mass squared $Q^2= m_{DD\pi}^2$, $S \to D^\ast D \to D D \pi$, and allow for the $D^\ast D$ pairs to rescatter, with the aim of describing the $\Tccp$ state with a $D^\ast D$ coupled channel $T$-matrix.

We start by discussing the coupled $T$-matrix for the $\cho$, $\cht$ channels (that we will often refer as $1$ and $2$, respectively), in which the $T_{cc}^+$ should show up as a pole. We write the $S$-wave $T$-matrix as:
\begin{equation}
T^{-1}(E) = V^{-1}(E) - \mathcal{G}(E)~,
\end{equation}
where the diagonal matrix $\mathcal{G}(E)$ contains the loop functions, to be discussed below, and the interaction kernels are contained in the matrix $V(E)$.  Both channels have $I_z=0$, and their isospin decomposition reads:
\begin{align}
\ket{D^{\ast+}D^0} & = - \frac{1}{\sqrt{2}} \left( \ket{D^\ast D,I=1} + \ket{D^\ast D,I=0} \right)~,\\
\ket{D^{\ast0}D^+} & = - \frac{1}{\sqrt{2}} \left( \ket{D^\ast D,I=1} - \ket{D^\ast D,I=0} \right)~,
\end{align}
and hence the interaction kernels can be written in terms of two isospin components, $C_{I=0,1}$, as:
\begin{equation}\label{eq:Vmat}
V(E) = \frac{1}{2} \left(\begin{array}{cc} 
C_0 + C_1 & C_1 - C_0 \\
C_1 - C_0 & C_0 + C_1
\end{array}\right)~.
\end{equation}
We take these two amplitudes $C_{0,1}$ to be constant, as a first approximation. A similar interaction is discussed in Ref.~\cite{Fleming:2021wmk}. The $\mathcal{G}$-matrix elements are the $\cho$ and $\cht$ loop functions, regularized by means of a Gaussian cutoff:
\begin{equation}\label{eq:Gloop}
G_i(E) = \int \frac{\mathrm{d}^3 \vec{k}}{(2\pi)^3} \frac{e^{-\frac{2\vec{k}^2}{\Lambda^2}}}{E-E_\text{th}^i - \frac{\vec{k}^2}{2\mu_i}}~,
\end{equation}
where $E_\text{th}^i$ and $\mu_i$ are respectively the thresold and the reduced mass of the channel. To take into account the width of the $D^\ast$, the loop functions are computed and then analytically continued to complex values of the $D^\ast$ mass, $m_{D^\ast} \to m_{D^\ast} - i \Gamma_{D^\ast}/2$. We take two values for the cutoff, $\Lambda=0.5\ \text{GeV}$ and $\Lambda=1.0\ \text{GeV}$. The $V$-matrix elements depend now on the cutoff, $C_I(\Lambda)$, and do not have specific meaning without specifying the cutoff.

\begin{figure}[t]\centering
$\DN \pi^+ \DN \to$%
\ifthenelse{\boolean{CompileDiagrams}}{\tikzsetnextfilename{DiagramsA1}
\DDpiDiagramTree{$\alpha$}{$\DsC$}{$\pi^+$}{$\DN$}{$\DN$}}{\raisebox{-0.33\height}{\includegraphics{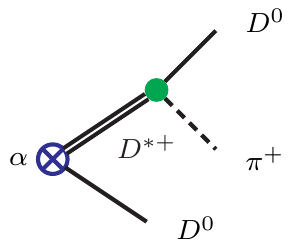}}}
$+$
\ifthenelse{\boolean{CompileDiagrams}}{\tikzsetnextfilename{DiagramsA3}
\DDpiDiagramLoop{$\alpha$}{$\DsC$}{$\pi^+$}{$\DN$}{$\DN$}{$\DsC$}{$\DN$}}{\raisebox{-0.33\height}{\includegraphics{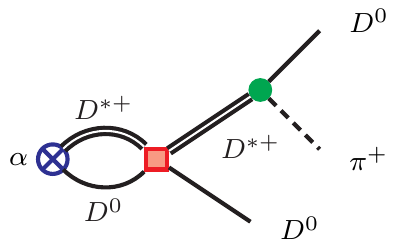}}}
$+$
\ifthenelse{\boolean{CompileDiagrams}}{\tikzsetnextfilename{DiagramsA2}
\DDpiDiagramLoop{$\beta$}{$\DsC$}{$\pi^+$}{$\DN$}{$\DN$}{$\DsN$}{$\DC$}}{\raisebox{-0.33\height}{\includegraphics{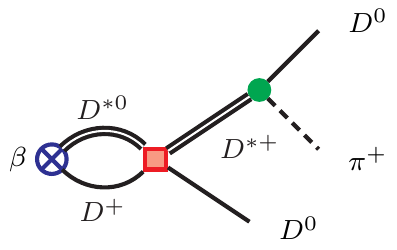}}}
\caption{Diagrams contributing to the $\DN\,\DN\,\pi^+$ final state production. Shown here are the $t$-channel diagrams, the $u$-channel ones are obtained by changing $\DN(p_1) \leftrightarrow \DN(p_2)$. Blue, cross vertices: coupling of the source to $D^\ast D$ states. Red, square vertices: $D^\ast D \to D^\ast D$ scattering amplitudes. Green, circle vertices: $D^\ast \to D \pi$ amplitudes. \label{fig:diagramsDZDZpiP}}
\end{figure}

The final state in the spectrum reported by the \LHCb collaboration in Ref.~\cite{LHCb:2021vvq} is $\DN\DN\pi^+$. With an eye to predict the spectra of other $DD\pi$ final states, we write the amplitude for the $DD\pi$ production from the decay $S(Q) \to D(p_1) D(p_2) \pi(p_\pi)$ in a generic way as:
\begin{equation}\label{eq:GenericAmplitudeDDpi}
\mathcal{M}_\lambda(Q^2,s,t,u) = g_{D^\ast D \pi}\, p_\pi^\nu \epsilon_S^{\mu}(\lambda) \left[ %
\frac{K_t(Q^2)}{t-\mDsSt} \left( - g_{\mu\nu} + \frac{k^{(t)}_\mu k^{(t)}_\nu}{t} \right)
+
\frac{K_u(Q^2)}{u-\mDsSu} \left( - g_{\mu\nu} + \frac{k^{(u)}_\mu k^{(u)}_\nu}{u} \right)
\right]~.
\end{equation}
Here, $\epsilon_S{(\lambda)}$ is the polarization vector of the axial source $S$. The Mandelstam variables are defined through:
\begin{subequations}\begin{align}
s & = (p_1+p_2)^2~,\\
t & = (p_1 + p_\pi)^2~,\\
u & = (p_2 + p_\pi)^2~,
\end{align}
and it holds that:
\begin{equation}
s + t + u = Q^2 + p_1^2 + p_2^2 + p_{\pi}^2~.
\end{equation}
\end{subequations}
We note that the masses of the external $D$ and $\pi$ mesons can be different for different $DD\pi$ final states. In Eq.~\eqref{eq:GenericAmplitudeDDpi} we have factored out the $D^\ast \to D \pi$ coupling constant $g_{D^\ast D \pi}$ as well as the pion momentum that arises from this vertex,
\begin{equation}
t^\mu_{D^\ast \to D\pi} \propto g_{D^\ast D \pi} \, \CIsos{}{}{} p_{\pi}^\mu~.
\end{equation}
The isospin coefficients are:
\begin{equation}
\CIsos{+}{0}{+} = \CIsos{0}{+}{-} = -\sqrt{2}\CIsos{0}{0}{0} = \sqrt{2}\CIsos{+}{+}{0} = 1~.
\end{equation}
In the $D^\ast$ propagator we have taken into account the possibility that the mass of the propagating vector is different in the $t$ and $u$ channels, as it happens for instance in the $D^+ D^0 \pi^0$ final state production. The $K_{t,u}(Q^2)$ functions are specific to each of the $DD\pi$ final states considered, and they contain the particular $D^\ast D$ dynamics associated to that final state. For instance, as shown in the diagrams of Fig.~\ref{fig:diagramsDZDZpiP}, for the final state $\DN\DN \pi^+$ we would have:
\begin{equation}
K_t(Q^2) = \alpha \left( 1 + G_{1}(Q^2) T_{11}(Q^2) \right) C_{\DsC \to \DN \pi^+} + \beta G_{2}(Q^2) T_{12}(Q^2)\,  C_{\DsC \to \DN \pi^+}~.
\end{equation}
Because of the symmetry in the $DD$ state, we have $K_u(Q^2) = K_t(Q^2)$ for the $\DN\DN\pi^+$ final state as well as for the $\DC\DC\pi^0$. Above, $\alpha$ and $\beta$ are the unknown strength of the $\DsC\DN$ and $\DsN\DC$ production vertices, respectively. As an approximation, these are taken as constant, similarly as done in Ref.~\cite{Dai:2021wxi}. This is a key assumption of our work, although again it seems a reasonable one given the small window of energies that we aim to describe. It is also important to note that the functions $K_{t,u}(Q^2)$ contain both the $D^\ast D$ rescattering through the loop terms, that will give rise to resonant contributions, if any, but also the tree level mechanisms, which can act as a background to the $DD\pi$ spectra. The $K_{t,u}(Q^2)$ functions for other $DD\pi$ final state are given in Appendix~\ref{app:functions}.

The $DD\pi$ event distribution as a function of the $DD\pi$ invariant mass is computed as:
\begin{equation}\label{eq:NeVQ2_notconvoluted}
\Nevnotconv(Q^2) = \mathcal{N}_0 \left( \frac{Q_\text{th}^2}{Q^2} \right)^\frac{3}{2} 
\int_{s_\text{th}}^{s_\text{max}(Q^2)} \mathrm{d}s 
\int_{t_-(s,Q^2)}^{t_+(s,Q^2)} \mathrm{d}t 
\sum_{\lambda} \left\lvert \mathcal{M}_\lambda(Q^2,s,t,u) \right\rvert^2~,
\end{equation}
where $\mathcal{N}_0$ is an unknown normalization constant (to be fitted), $\sqrt{Q^2_\text{th}}$ and $\sqrt{s_\text{th}}$ are the appropriate $DD\pi$ and $DD$ thresholds, respectively, $s_\text{max}(Q^2) = (\sqrt{Q^2} - m_\pi)^2$ is the upper limit for the $DD$ system invariant mass, and $t_\pm(s,Q^2)$ are the usual limits in the $t$ variable (see {\it e.g.} Ref.~\cite{ParticleDataGroup:2020ssz}). We not that there is an additional $1/2$ symmetry factor not shown in Eq.~\eqref{eq:NeVQ2_notconvoluted} but included in the calculations for the final states containing a $D^0D^0$ or $D^+D^+$ pair. The integrand reads (absorbing $g_{D^{\ast} D \pi}$ in the overall normalization constant):
\begin{align}
\sum_{\lambda} \left\lvert \mathcal{M}_\lambda \right\rvert^2 & = 
\left\lvert K_t(Q^2) \right\rvert^2 \frac{H_t(Q^2,s,t,u)}{\left \lvert t - \mDsSt \right\rvert^2} +
\left\lvert K_u(Q^2) \right\rvert^2 \frac{H_u(Q^2,s,t,u)}{\left \lvert u - \mDsSu \right\rvert^2} \nonumber\\
& 
+ 2\, \text{Re}\, K_t(Q^2) {K_u(Q^2)}^\ast \frac{H_{tu}(Q^2,s,t,u)}{(t- \mDsSt)(u-{\mDsSu})^\ast}
~,
\end{align}
and it can be seen that the products of the $K_{t,u}(Q^2)$ functions factor out of the integrals. The functions $H_{t,u,tu}(Q^2,s,t,u)$ are kinematical, and arise from the contractions with the polarization vector of the source $S$ upon summation over the latter. Due to the narrow width of the $D^\ast$, the integrals of the propagator times these kinematical functions are quasi-two body phase space functions \cite{LHCb:2021auc}.

\section{Results}\label{sec:results}
\subsection{Fit}\label{subsec:fits}
\begin{figure}[t]\centering
\ifthenelse{\boolean{CompileFigures}}{\input{PlotData.tex}}{\includegraphics{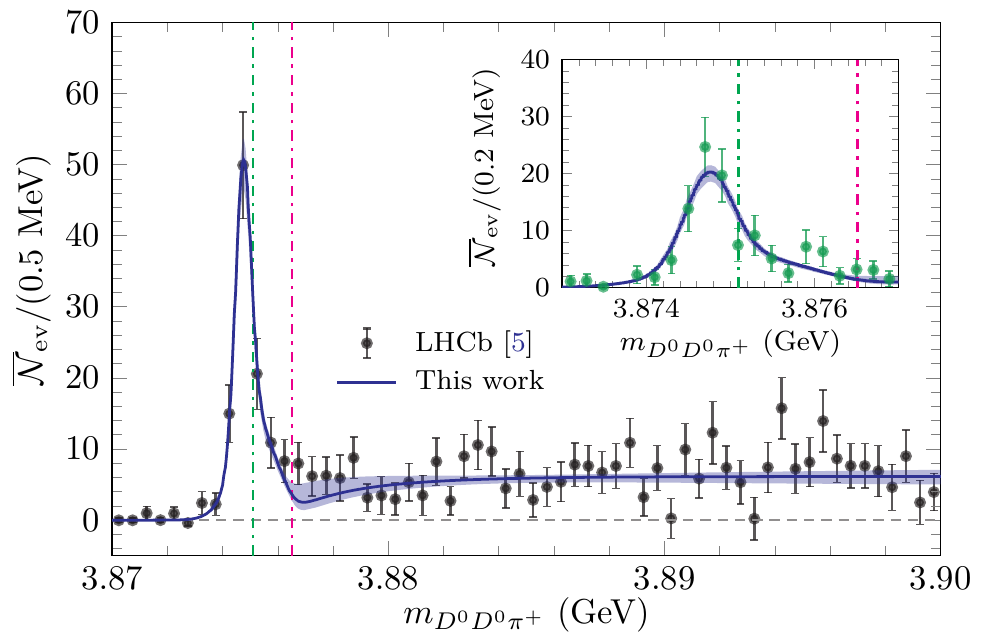}}
\caption{$\DN\DN\pi^+$ spectrum showing the prominent $\Tccp$ signal. The data come from the \LHCb collaboration \cite{LHCb:2021vvq}, and the theoretical curve is calculated with Eq.~\eqref{eq:MeVQ2_convoluted}, and the parameters of Table~\ref{table:parameters}. The uncertainty is computed through MC bootstrap with resampling of the data, as explained in the text. The inset shows in greater detail the energy region around the peak. The green and purple vertical dashed-dotted lines represent the $\cho$ and $\cht$ thresholds, respectively.\label{fig:data}}
\end{figure}
\begin{figure}[t]\centering
\ifthenelse{\boolean{CompileFigures}}{\input{PullPlots200500.tex}}{\includegraphics{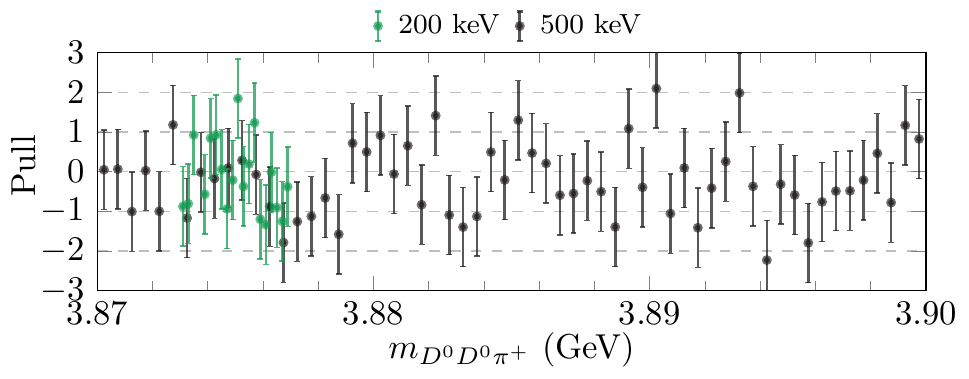}}
\caption{Pull of the theoretical curve with respect to the data for the central fit shown in Fig.~\ref{fig:data}. The black (green) points corresponds with the $500\ \text{keV}$ ($200\ \text{keV}$) bins.\label{fig:PullCentralFit}}
\end{figure}
We aim to compare the event distribution $\Nevnotconv(Q^2)$ in Eq.~\eqref{eq:NeVQ2_notconvoluted} with the experimental one obtained by the \LHCb collaboration \cite{LHCb:2021vvq}. Because of the global normalization, we fix $\alpha=1$ and the free parameters are thus $C_0$, $C_1$, $\beta$, and $\mathcal{N}_0$. To take into account the experimental resolution $\delta \simeq 400\ \text{keV}$, we convolute Eq.~\eqref{eq:NeVQ2_notconvoluted} with an energy resolution function \cite{LHCb:2021vvq,LHCb:2021auc},\footnote{The resolution function reads:
\begin{equation*}
R_\text{\LHCb}(E,E') = \alpha \mathcal{G}(E,E',\sigma_1) + (1-\alpha)\mathcal{G}(E,E',\sigma_2)~,
\end{equation*}
where $\mathcal{G}(x,\mu,\sigma)$ is a standard gaussian distribution, and the parameters are $\sigma_1 = 1.05 \times 263\ \text{keV}$, $\sigma_2 = 2.413\sigma_1$, and $\alpha = 0.778$, taken from Ref.~\cite{LHCb:2021auc}.}
\begin{equation}\label{eq:MeVQ2_convoluted}
\Nevconv(Q^2) = \int \mathrm{d}E\, R_\text{\LHCb}\left(E,\sqrt{Q^2}\right)\, \Nevnotconv(E^2)~.
\end{equation}
We fit the four free parameters to the available $79$ experimental points ($60$ points in energy bins of $500\ \text{keV}$, and $19$ in energy bins of $200\ \text{keV}$, as can be seen in Fig.~\ref{fig:data}), obtaining the values in Table~\ref{table:parameters}. The $\chi^2$ of the best fit is $\chi^2/\text{dof} = 71.1/(60+19-4)=0.95$ ($\Lambda=1\ \text{GeV})$ and $70.6/(60+19-4)=0.92$ ($\Lambda=0.5\ \text{GeV}$), and the good agreement with the data can be seen in Fig.~\ref{fig:data}, where the blue, solid line represents our best fit. The uncertainties are obtained by bootstrapping the fits with Montecarlo (MC) resampling of the data (with $\sim 2000$ MC steps) and thus the correlations are taken into account. As a sanity check, the errors of the parameters are similar to those obtained with \textsc{Minuit} \cite{James:1994vla} with both \texttt{minimize} and \texttt{minos}. Also, as shown in Fig.~\ref{fig:PullCentralFit}, the pull of the theoretical curve with respect to the data seems randomly distributed.\footnote{For each data point, the pull is defined as the difference between the theoretical curve and the experimental point divided by the error of the latter.}

\begin{table}[t]\centering\small%
\begin{tabular}{ccc}%
Parameter                 & $\Lambda = 1.0\ \text{GeV}$    & $\Lambda=0.5\ \text{GeV}$ \\ \hline\hline
$C_0(\Lambda)\,[ \text{fm}^2 ] $        & $-0.7008(22)$    & $-1.5417(121)$ \\
$C_1(\Lambda)\,[ \text{fm}^2 ] $        & $-0.440(79)$     & $-0.71(27)$ \\
$\beta/\alpha$                          & $ 0.228(108)$    & $0.093(79)$ \\
$\chi^2/\text{dof}$                     & $ 0.95$          & $0.92$ \\ \hline
\end{tabular}
\caption{Value of the parameters from the fit of the $\DN\DN\pi^+$ event distribution, Eq.~\eqref{eq:MeVQ2_convoluted}, to the \LHCb experimental results \cite{LHCb:2021vvq}. As discussed in Subsec.~\ref{subsec:degeneracy}, this is the \textit{isoscalar} solution. The \textit{isovector} solution, degenerated with the \textit{isoscalar} one, is obtained by the simultaneous replacement $C_0 \leftrightarrow C_1$ and $\beta \leftrightarrow -\beta$.\label{table:parameters}}
\end{table}

Our curve shares some features with the \LHCb fit \cite{LHCb:2021vvq}, like the prominent $\Tccp$ signal (obviously), a sort of small dip around the $\cht$ threshold, and a phase-space-like background opening around the $\cho$ threshold. We note that in this work these features stem from the shape of the $T$-matrix and the production mechanism. Our curve lies a bit higher (and so closer to the data) than the \LHCb one at the peak. Near the peak, our curve lies less than $1\sigma$ away from most of the experimental points, also in the zoomed energy region, as can be seen by the pull of the data in Fig.~\ref{fig:PullCentralFit}.

\subsection{Fit degeneracy}\label{subsec:degeneracy}
We note that under the simultaneous exchange $\beta \to -\beta$ and $C_0 \leftrightarrow C_1$ the $K_{t,u}(Q^2)$ functions for the $\DN\DN\pi^+$ final state are unchanged. Thus, for any fit that we find with specific values of the free parameters, there is another one with the same $\chi^2$ and $\Nevconv(Q^2)$, {\it e.g.} Fig.~\ref{fig:data} remains unchanged. Physically speaking, this reflects the $I_z=0$ nature of this specific $D^\ast D$ and $DD\pi$ final state, and our inability to distinguish the isospin ($I=0$ or $I=1$) from the $\DN\DN\pi^+$ spectra {\it alone}. We will refer to our main fit as {\it isoscalar solution}, and the additional one as the {\it isovector} one. This {\it symmetry} also affects the $I_z=0$ $\DC\DC\pi^-$ and $\DN\DC\pi^0$ spectra,\footnote{It induces a change in the sign of both the $K_t(Q^2)$ and $K_u(Q^2)$ functions, so that the spectra remain the same.} and hence each of the three spectra will be equal in both solutions. However, the spectra for the final states with $I_z = \pm 1$ ($\DC\DC\pi^0$, $\DN\DC\pi^+$, $\DC\DN\pi^-$, and $\DN\DN\pi^0$, stemming from either $\DsC\DC$ or $\DsN\DN$), being these purely isovector, do not suffer from this degeneracy in the solution. These spectra, as we will show below, should allow to distinguish the $\Tccp$ isospin.

We remark here that, although the $\DN\DN\pi^+$ spectrum alone does not allow to determine the $\Tccp$ isospin, the \LHCb collaboration has given arguments in favour of its interpretation as an isoscalar state. In particular, in Ref.~\cite{LHCb:2021auc} the experimental spectrum for the $\DN\DC\pi^+$ final state is shown, although with much less statistics than the $\DN\DN\pi^+$ case. Since there is no enhancement near the $\DsC\DC$ threshold, this points to the isoscalar interpretation of $\Tccp$. For completeness, though, and in await of more definitive spectra, we show and discuss in what follows both the \textit{isoscalar} and the \textit{isovector} solutions, bearing in mind that the isoscalar one is more favoured.

\subsection{Pole position, couplings, scattering length}\label{subsec:poles}
\begin{figure}[t]\centering
\ifthenelse{\boolean{CompileFigures}}{\input{PlotDet1mVG.tex}}{\includegraphics{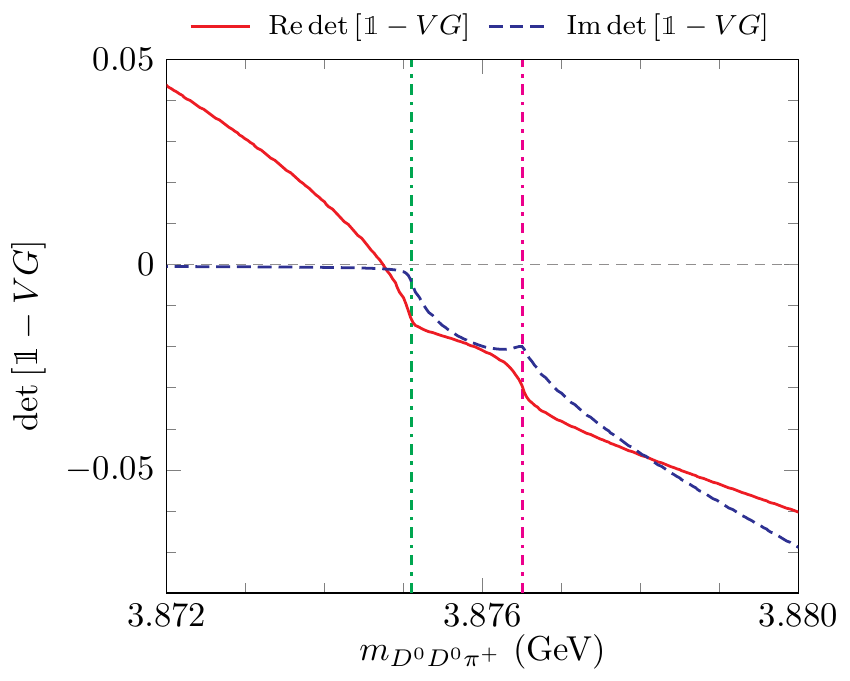}}
\caption{Real (red solid line) and imaginary (blue dashed line) parts of $\det \left[ \mathbbm{1} - VG \right]$. The vertical dashed-dotted line represent the $\cho$ (green) and $\cht$ (magenta) thresholds. It can be observed that the real part has a zero very close to the $\Tccp$ pole.\label{fig:PlotDet}}
\end{figure}

\begin{figure}[t]\centering
\ifthenelse{\boolean{CompileFigures}}{\input{figNotConvolutedSpectrum.tex}}{\includegraphics{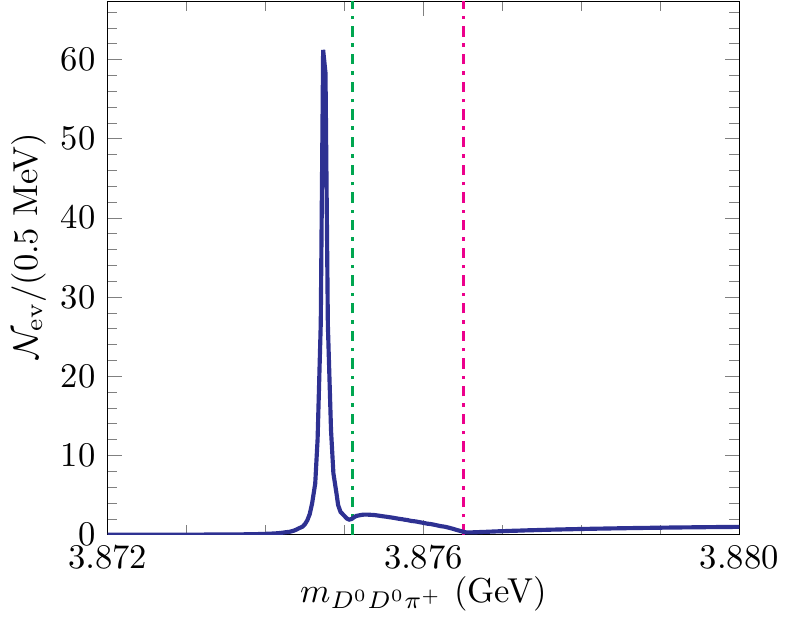}}
\caption{The $\DN\DN\pi^+$ spectrum \textit{before} convoluting with the resolution function, as a mass of the $\DN\DN\pi^+$ invariant mass. The very narrow peak corresponds to the $\Tccp$ bound state. The vertical dashed-dotted line represent the $\cho$ (green) and $\cht$ (magenta) thresholds. \label{fig:NotConvolutedSpectrum}}
\end{figure}

The prominent $\Tccp$ signal should appear as a bound state pole in the $\cho$, $\cht$ coupled channel $T$-matrix,\footnote{The additional factors in Eq.~\eqref{eq:TMatPole} are included for normalization purposes}
\begin{equation}\label{eq:TMatPole}
\sqrt{2m_{D^\ast_i}m_{D_i}} \sqrt{2m_{D^\ast_j}m_{D_j}} T_{ij}(E) = \frac{g_i g_j}{E^2 - \left( M_{\Tccp} - i \displaystyle\frac{\Gamma_{\Tccp}}{2} \right)^2} + \cdots~,
\end{equation}
and thus as a zero of $\det \left( \mathbbm{1} - V\,G \right)$. In Fig.~\ref{fig:PlotDet} we show this determinant as a function of the $m_{\DN\DN\pi^+}$ invariant mass. It can be seen that the real part goes through zero close and below the  $\cho$ threshold. However, the imaginary part is not zero because of the finite width of the $D^\ast$. For this reason, the pole is indeed not located on the real axis but on the complex plane, and we find:
\begin{subequations}\label{eq:PolePosition}
\begin{align}
\Delta M_{\Tccp} = M_{\Tccp} - m_{\DsC} - m_{\DN} 
& = -357(29)\ \text{keV}~,
& \Gamma_{\Tccp} 
& = 77(1)\ \text{keV} & [\Lambda=1.0\ \text{GeV}]~,\\
& = -356(29)\ \text{keV}~,\,
& 
& = 78(1)\ \text{keV} & [\Lambda=0.5\ \text{GeV}]~. \label{eq:PolePosition500}
\end{align}
\end{subequations}
Due to the simple parameterization used for the interaction [\textit{cf.} Eq.~\eqref{eq:Vmat}], the model has freedom to fix the mass of the bound state, but not its width, which in our approach stems essentially from the widths of the $D^\ast$ mesons, $\Gamma_{\DsC} = 83.4\ \text{keV}$ and $\Gamma_{\DsN}=56.2\ \text{keV}$ (and it is close to the first of them). When compared with the BW determination by the LHCb collaboration in Ref.~\cite{LHCb:2021vvq} [\textit{cf.} Eq.~\eqref{eq:LHCbBW}], the value for the masses lie within $1-2\sigma$ deviation, although our width is much smaller than the reported BW one, $\Gamma_\text{BW} = 410(165)\ \text{keV}$. Indeed, the bound state appears very narrow in the original (\textit{i.e.} not convoluted) spectrum $\Nevnotconv(m_{\DN\DN\pi^+}^2)$, as shown in Fig.~\ref{fig:NotConvolutedSpectrum}. This not-convoluted spectrum resembles the ones obtained by Ref.~\cite{LHCb:2021auc} (see Extended Data Fig.~8 therein) and Ref.~\cite{Feijoo:2021ppq}. On the other hand, in the improved analysis of Ref.~\cite{LHCb:2021auc} by the \LHCb collaboration [\textit{cf.} Eq.~\eqref{eq:LHCbPole}], an amplitude model is studied that presents a pole at $\Delta M = -360(40)\ \text{keV}$ and $\Gamma = 48(2)\ \text{keV}$, being the agreement of our calculation [Eq.~\eqref{eq:PolePosition}] much better with this result. Still, our width is larger than that of Ref.~\cite{LHCb:2021auc}, so more investigation will be necessary. We point here that our calculation only includes hadronic channels, and thus the width can suffer from systematic deviations.

For the couplings, as defined in Eq.~\eqref{eq:TMatPole}, we obtain (the imaginary part is negligible):
\begin{subequations}\label{eq:Couplings}\begin{align}
\hphantom{\pm} g_{\cho} 
& = 4.13(12)\ \text{GeV}~, &  \pm g_{\cht} & = 3.53(33)\ \text{GeV}\quad [\Lambda=1.0\ \text{GeV}]~,\\
& = 4.36(16)\ \text{GeV}~, &               & = 3.67(50)\ \text{GeV}\quad [\Lambda=0.5\ \text{GeV}]~.
\end{align}
\end{subequations}
The sign of $g_{\cht}$ is negative (positive) in the isoscalar (isovector) solution. In the exact isospin limit, one would have $g_1 = - g_2$ ($g_1 = +g_2$) for an isoscalar (isovector) state (see {\it e.g.} Ref.~\cite{Fleming:2021wmk}). The fact that both couplings are similar indicates that isospin symmetry is not very broken at the level of interactions, despite the gap between thresholds. The values obtained here are similar to those in Ref.~\cite{Feijoo:2021ppq}, although the difference in the couplings of both channels is larger in our calculation.

We can also compute the scattering length of the $\cho$ channel, which in our normalization is given by:
\begin{equation}
a_{\cho} = - \frac{\mu_1}{2\pi}T_{11}(E^{1}_\text{th})~,
\end{equation}
and we get:
\begin{subequations}\begin{align}
a_{\cho}
& = -7.99(46) + i\,2.21(28)\ \text{fm} \quad [\Lambda=1.0\ \text{GeV}]~,\\
& = -8.56(49) + i\,2.61(32)\ \text{fm} \quad [\Lambda=0.5\ \text{GeV}]~.
\end{align}\end{subequations}
The imaginary part does not vanish because the $\cho$ system decays to $DD\pi$ (which in our approach is taken into account by the presence of the $D^\ast$ widths), and it is sizeable because the width of the $\Tccp$, though small, is not negligible when compared to its binding energy. These values are similar to those obtained in Ref.~\cite{LHCb:2021auc}.

\subsection{Molecular state?}\label{subsec:molecular}
The Weinberg compositeness criterium \cite{Weinberg:1965zz} is often used to assess the ``molecularness'' of a given state, and there is quite a lot of activity in this topic (see \textit{e.g.} Refs.~\cite{Baru:2003qq,Gamermann:2009uq,Yamagata-Sekihara:2010kpd,Hyodo:2011qc,Aceti:2012dd,Guo:2015daa,Sekihara:2014kya,Oller:2017alp,Matuschek:2020gqe} and references therein). A generalization to coupled channels performed in Ref.~\cite{Gamermann:2009uq} allows to interpret the quantities $P_i$,
\begin{equation}\label{eq:pigi2G}
P_i = -g_i^2 \frac{\mathrm{d}G(E^2)}{\mathrm{d} E^2}~,
\end{equation}
as the probabilities of finding the bound state in a given channel. We find the values:
\begin{subequations}\label{eq:Probabilities}\begin{align}
\hphantom{\pm} P_{\cho} 
& = 0.77(4)~, &  P_{\cht} & = 0.23(4)~,& [\Lambda=1.0\ \text{GeV}]~,\\
& = 0.79(5)~, &           & = 0.21(5)~,& [\Lambda=0.5\ \text{GeV}]~, \label{eq:Probabilities500}
\end{align}
\end{subequations}
which turn out to be remarkably independent of the cutoff $\Lambda$. In the exact isospin limit, one can check that the probabilities are $P_{\cho} = P_{\cht} = 0.5$, or, in terms of definite isospin states, the molecular probability is exactly one, $P_I = 1$ (regardless of the isospin $I=0$ or $I=1$ of the state, see Subsec.~\ref{subsec:degeneracy}). However, it must be said that the fact that $P_I=1$ (or equivalently $P_{\cho} + P_{\cht}=1$) is strictly built in the model and in Eq.~\eqref{eq:pigi2G} due to the simple, constant parameterization of the $V$ matrix [\textit{cf.} Eq.~\eqref{eq:Vmat}]. It is not a consequence of the specific values of the constants $C_I(\Lambda)$.

We next discuss the results that are obtained when the original Weinberg arguments \cite{Weinberg:1965zz} (see also Ref.~\cite{Matuschek:2020gqe}) about the compositeness are used. Certainly, the $\Tccp$ bound state that we obtain seems to satisfy the three applicability requirements (stable or very narrow state, coupling to a two-channel threshold not much above the location of the state, and $S$-wave) of Ref.~\cite{Weinberg:1965zz}. The formula for the molecular probability reads:
\begin{equation}\label{eq:PIWeinIsLim}
P_{I} = \sqrt{\frac{1}{1 + \left\lvert \displaystyle\frac{2r_I}{a_I} \right\rvert}}~.
\end{equation}
For simplicity, we discuss the results in the isospin\footnote{Numerically, we take $m_{D} = (m_{D^{+}} + m_{D^{0}})/2$, and similarly for the vector mesons.} and zero $D^\ast$ width limits, where the concept of molecule should make more sense. In the isospin limit, the $\cho$ and $\cht$ $T$-matrix can be diagonalized into $I=0$ and $I=1$ amplitudes, $T_I(E) = C_I^{-1} - G(E)$. The $\Tccp$ is in this case a bound state with zero width too, and the scattering length and effective range, that we denote now $a_I$ and $r^0_I$, are also real. We find:\footnote{Incidentally, we note that the results for the $D^\ast D$ scattering length and the effective range in the isospin limit are similar to those of nucleon-nucleon scattering in the deuteron channel, $a_\text{d} = -5.41\ \text{fm}$ and $r_\text{d} = +1.75\ \text{fm}$ \cite{Weinberg:1965zz}.}
\begin{subequations}\begin{align}
a_{I} & = -5.18(16)\ \text{fm}~, & r_I & = 0.63\ \text{fm}~, & P_I & = 0.897(3)~,& [\Lambda=1.0\ \text{GeV}]~,\\
      & = -5.57(25)\ \text{fm}~, &     & = 1.26\ \text{fm}~, &     & = 0.830(5)~,& [\Lambda=0.5\ \text{GeV}]~.
 \label{eq:ResultsPIWeinIsLim}
\end{align}\end{subequations}
We see that a large molecular component between $80\%$--$90\%$ is obtained. Because of the constant value taken for the kernels $V_I = C_I(\Lambda)$, the effective range $r$ is solely determined by the cutoff, and does not depend on the $C_I(\Lambda)$ values, which can affect the determination of the molecular probability. That being said, in Appendix~\ref{app:vpotGeneralization} we have investigated more complex parameterizations of the matrix elements of $V$, that would allow better determinations of the effective range $r$. While these alternative parameterizations improve slightly the quality of the fit, they do not change dramatically this probability. The molecular probability is always large, and thus our conclusion is that the molecular description fits nicely for the $\Tccp$.

\section{Predictions}\label{sec:otherpredictions}
\subsection[Predictions of the spectrum for different $DD\pi$ final states]{\boldmath Predictions of the spectrum for different $DD\pi$ final states}\label{subsec:PredictionsSpectra}
As previously mentioned, the $\DN\DN\pi^+$ spectrum alone does not allow to determine the $\Tccp$ isospin. Instead, other $DD\pi$ spectra are needed, and hence we make a prediction of such additional spectra here, independently of the nature of the $\Tccp$ state.

\begin{figure}\centering
$\DC \pi^- \DC \to$%
\ifthenelse{\boolean{CompileDiagrams}}{\tikzsetnextfilename{DiagramsC1}
\DDpiDiagramTree{$\beta$}{$\DsN$}{$\pi^-$}{$\DC$}{$\DC$}}{\raisebox{-0.33\height}{\includegraphics{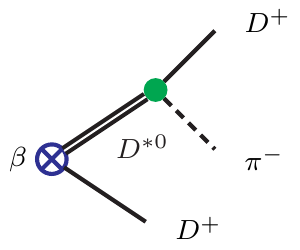}}}
$+$
\ifthenelse{\boolean{CompileDiagrams}}{\tikzsetnextfilename{DiagramsC2}
\DDpiDiagramLoop{$\beta$}{$\DsN$}{$\pi^-$}{$\DC$}{$\DC$}{$\DsN$}{$\DC$}}{\raisebox{-0.33\height}{\includegraphics{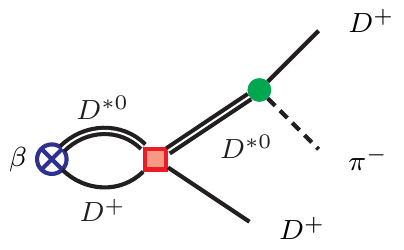}}}
$+$
\ifthenelse{\boolean{CompileDiagrams}}{\tikzsetnextfilename{DiagramsC3}
\DDpiDiagramLoop{$\alpha$}{$\DsN$}{$\pi^-$}{$\DC$}{$\DC$}{$\DsC$}{$\DN$}}{\raisebox{-0.33\height}{\includegraphics{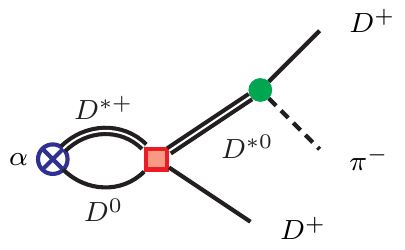}}}
\caption{Same as Fig.~\ref{fig:diagramsDZDZpiP}, but for the $\DC\,\DC\,\pi^-$ final state production.\label{fig:diagramsDPDPpiM}}
\end{figure}

\begin{figure}[t]
$\DN \pi^0 \DC \to$%
\ifthenelse{\boolean{CompileDiagrams}}{\tikzsetnextfilename{DiagramsD1}
\DDpiDiagramTree{$\beta$}{$\DsN$}{$\pi^0$}{$\DN$}{$\DC$}}{\raisebox{-0.33\height}{\includegraphics{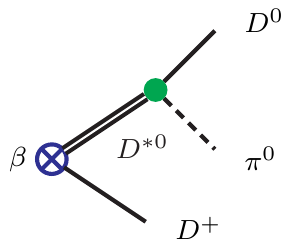}}}
$+$
\ifthenelse{\boolean{CompileDiagrams}}{\tikzsetnextfilename{DiagramsD2}
\DDpiDiagramLoop{$\beta$}{$\DsN$}{$\pi^0$}{$\DN$}{$\DC$}{$\DsN$}{$\DC$}}{\raisebox{-0.33\height}{\includegraphics{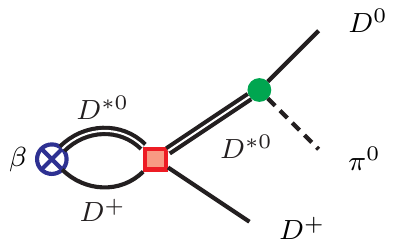}}}
$+$
\ifthenelse{\boolean{CompileDiagrams}}{\tikzsetnextfilename{DiagramsD5}
\DDpiDiagramLoop{$\alpha$}{$\DsN$}{$\pi^0$}{$\DN$}{$\DC$}{$\DsC$}{$\DN$}}{\raisebox{-0.33\height}{\includegraphics{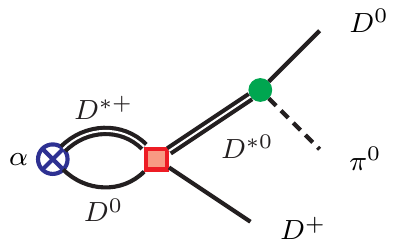}}}
\\\hphantom{$\DN \pi^0 \DC$}%
$+$
\ifthenelse{\boolean{CompileDiagrams}}{\tikzsetnextfilename{DiagramsD4}
\DDpiDiagramTree{$\alpha$}{$\DsC$}{$\pi^0$}{$\DC$}{$\DN$}}{\raisebox{-0.33\height}{\includegraphics{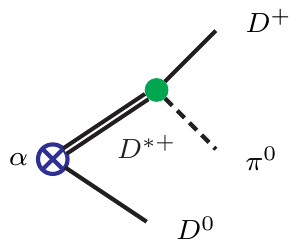}}}
$+$
\ifthenelse{\boolean{CompileDiagrams}}{\tikzsetnextfilename{DiagramsD6}
\DDpiDiagramLoop{$\alpha$}{$\DsC$}{$\pi^0$}{$\DC$}{$\DN$}{$\DsC$}{$\DN$}}{\raisebox{-0.33\height}{\includegraphics{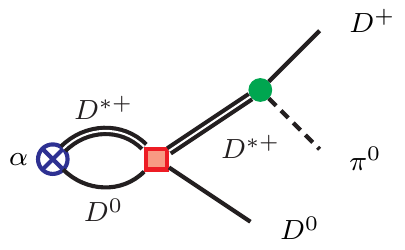}}}
$+$
\ifthenelse{\boolean{CompileDiagrams}}{\tikzsetnextfilename{DiagramsD3}
\DDpiDiagramLoop{$\beta$}{$\DsC$}{$\pi^0$}{$\DC$}{$\DN$}{$\DsN$}{$\DC$}}{\raisebox{-0.33\height}{\includegraphics{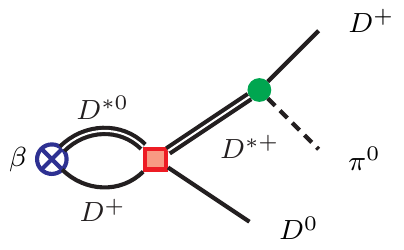}}}
\caption{Diagrams contributing to the $\DC\DN\pi^0$ production. $t$- and $u$-channels are explicitly shown.\label{fig:diagramsDZDPpiZ}}
\end{figure}

\begin{figure}[t]\centering
\ifthenelse{\boolean{CompileFigures}}{\input{PlotOtherSpectra_Iz0.tex}}{\includegraphics{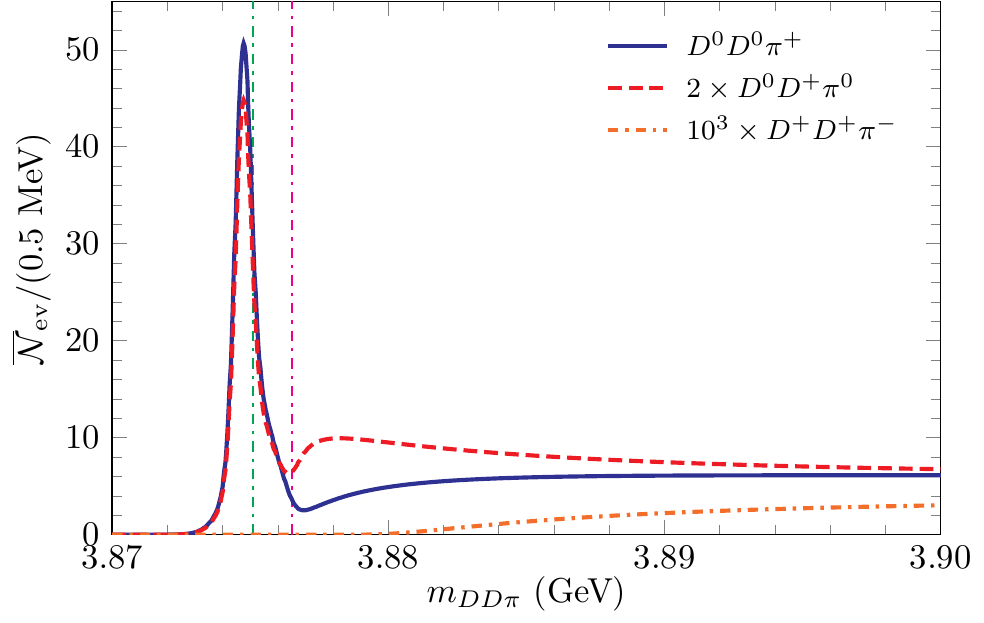}}
\caption{Predictions of the spectra for other $DD\pi$ final states with $I_z=0$. The same production mechanism and resolution as in the $\DN\DN\pi^+$ case (shown in Fig.~\ref{fig:data} and also here) are assumed. To facilitate the comparison, some of the spectra have been scaled. The vertical dashed-dotted line represent the $\cho$ (green) and $\cht$ (magenta) thresholds. \label{fig:OtherSpectraIz0}}
\end{figure}

The amplitudes for the final states $\DC\DC \pi^-$ and $\DN\DC\pi^0$, with $I_z = 0$ too, are diagramatically shown in Figs.~\ref{fig:diagramsDPDPpiM} and \ref{fig:diagramsDZDPpiZ}, respectively. Assuming the same production mechanisms and the same resolution as in the $\DN\DN\pi^+$ case, and ignoring possible differences in the reconstruction efficiencies of particles in the final states, we can predict the spectra for these final states. These are shown in Fig.~\ref{fig:OtherSpectraIz0}, and for the sake of comparison we have assumed the same normalization ({\textit i.e.} statistics).

The $\DC\DC\pi^-$ threshold lies $\sim 10\ \text{MeV}$ above the $\DN\DN\pi^+$ (and $\DC\DN\pi^0$) threshold, which makes a huge difference given the $30\ \text{MeV}$ spanned by the data and the narrowness of the $\Tccp$ signal. In addition, the decaying $D^\ast$ in this channel is always a $\DsN$, which has a smaller width than that of $\DsC$. All in all, these phase-space facts make it so that the number of events in the $\DC\DC\pi^-$ spectrum is negligible, and there is no trace of the $T_{cc}^+$, as can be seen in Fig.~\ref{fig:OtherSpectraIz0} (dashed-dotted orange line). On the other hand, this does not happen for the $\DC\DN\pi^0$ final state, and as shown in Fig.~\ref{fig:OtherSpectraIz0} (dashed red line) the spectrum is comparable to (but smaller than) that of the $\DN\DN\pi^+$ final state (solid blue line). Therefore, this channel could be used to confirm the existence of the $\Tccp$ state. As explained in Subsec.~\ref{subsec:degeneracy}, the predicted spectra for these final states, shown in Fig.~\ref{fig:OtherSpectraIz0}, are the same for both the \textit{isoscalar} and the \textit{isovector} solutions.

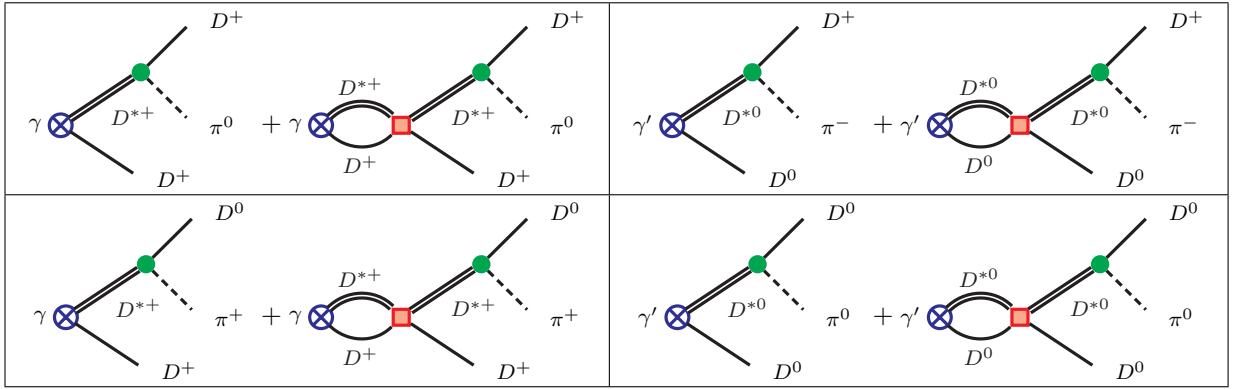
\begin{figure}[t]\centering
\input{DiagramsOtherIsovector.tex}
\caption{Diagrams contributing to final states with $I_z=\pm 1$. \label{fig:diagramsIzPM}}
\end{figure}

\begin{figure}[t]\centering
\ifthenelse{\boolean{CompileFigures}}{\input{PlotOtherSpectra_Iz1.tex}}{\includegraphics{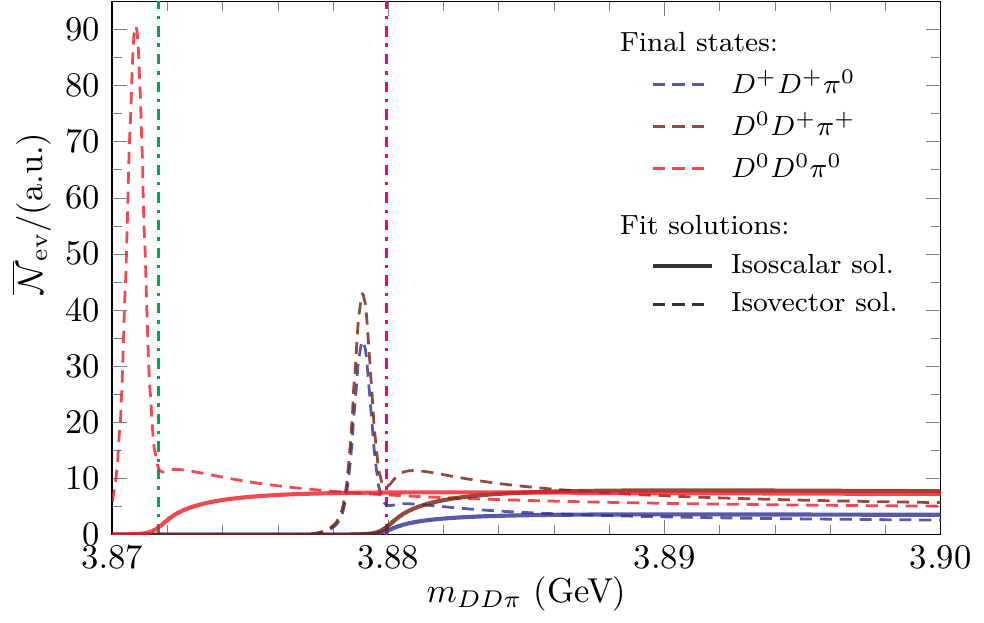}}
\caption{Prediction of the spectra for $DD\pi$ final states with $I_z = \pm 1$. The green and magenta vertical dashed-dotted lines represent the $\DsN\DN$ and $\DsC\DC$ thresholds, respectively. The thick solid (thin dashed) lines represent the spectra obtained when the isoscalar (isovector) solution is used. Different colors (blue, brown, and red) represent different final state ($\DC\DC\pi^0$, $\DN\DC\pi^+$, and $\DN\DN\pi^0$, respectively) spectra. The green and magenta vertical dashed-dotted lines mark the $\DsN\DN$ and $\DsC\DC$ thresholds, respectively. As explained in the text, the distribution for the $\DN\DC\pi^-$ final state is negligible and not shown here.\label{fig:OtherSpectraIz1}}
\end{figure}

We can also predict the spectrum for the $I_z = \pm 1$ final states, where the difference between both solutions should be visible. The diagrams used to compute these spectra are shown in Fig.~\ref{fig:diagramsIzPM}, and the line shapes (taking into account also here the resolution) are shown in Fig.~\ref{fig:OtherSpectraIz1}. Similarly as for $\DC\DC\pi^-$ case, the $\DN\DC\pi^-$ distribution is negligible due to the lack of phase space, and thus not shown in Fig.~\ref{fig:diagramsIzPM}.
 
The spectra predicted for the \textit{isoscalar} solution (solid thick lines in Fig.~\ref{fig:OtherSpectraIz1}) show no enhancement at threshold nor additional states, and are smooth phase-space distributions. In the \textit{isovector} solution, since the $\Tccp$ would have $I=1$, one thus expects the other two members of the isospin triplet, $T_{cc}^{++}$ and $T_{cc}^0$ to show up in these spectra, as seen indeed (dashed thin lines) in Fig.~\ref{fig:OtherSpectraIz1}. In this solution we find two additional poles with the following binding energies:

\begin{subequations}
\begin{align}
\Delta M_{\Tccpp} = M_{\Tccpp} - m_{\DsC} - m_{\DC} 
& = -873(53)\ \text{keV}
\quad [\Lambda=1.0\ \text{GeV}]~,\\
\hphantom{\Delta M_{\Tccz} = M_{\Tccz} - m_{\DsN} - m_{\DN} }
& = -842(67)\ \text{keV}
\quad [\Lambda=0.5\ \text{GeV}]~,
\end{align}\end{subequations}
\begin{subequations}\begin{align}
\Delta M_{\Tccz} = M_{\Tccz} - m_{\DsN} - m_{\DN} 
& = -838(52)\ \text{keV}
\quad [\Lambda=1.0\ \text{GeV}]~,\\
\hphantom{\Delta M_{\Tccpp} = M_{\Tccpp} - m_{\DsC} - m_{\DC} }
& = -825(67)\ \text{keV}
\quad [\Lambda=0.5\ \text{GeV}]~.
\end{align}
\end{subequations}
These states would be somewhat more bound than the $\Tccp$ because there is a single channel/threshold. Finally, we  must stress that the $\DC\DN\pi^+$ spectrum shown in Ref.~\cite{LHCb:2021auc} by the \LHCb collaboration, albeit with wide bins, agrees with the shape of the $\DC\DN\pi^+$ spectrum predicted in the \textit{isoscalar} solution (thick solid green line in Fig.~\ref{fig:OtherSpectraIz1}), and not with the \textit{isovector} one (thin dashed green line). As discussed in Subsec.~\ref{subsec:degeneracy}, this fact favours the $I=0$ interpretation of $\Tccp$ and thus our \textit{isoscalar} solution.

\subsection[Heavy Quark Spin Symmetry partners of $T_{cc}^+$]{\boldmath Heavy Quark Spin Symmetry partners of $T_{cc}^+$}\label{subsec:HQSS}
Heavy-Quark Spin Symmetry (HQSS) \cite{Neubert:1993mb,Manohar:2000dt} predicts that heavy-meson interactions are independent of the heavy-quark spin. Similar to the $D^{(\ast)}\bar{D}^{(\ast)}$ case \cite{Hidalgo-Duque:2012rqv,Guo:2013sya}, HQSS predicts that, up to $\mathcal{O}(1/M_Q)$, the interaction kernels of the $I(J^P)$ $D^\ast D^\ast$ systems are related to those of the $D^\ast D$ ones as:
\begin{subequations}\label{eqs:HQSSrel}
\begin{align}
\braket{D^\ast D^\ast,\, 0(1^{+})}{\hat{V}}{D^\ast D^\ast,\, 0(1^{+})} & = 
\braket{D^\ast D     ,\, 0(1^{+})}{\hat{V}}{D^\ast D     ,\, 0(1^{+})} = V_0~, \label{eq:HQSSI0}\\
\braket{D^\ast D^\ast,\, 1(2^{+})}{\hat{V}}{D^\ast D^\ast,\, 1(2^{+})} & = 
\braket{D^\ast D     ,\, 1(1^{+})}{\hat{V}}{D^\ast D     ,\, 1(1^{+})} = V_1~. \label{eq:HQSSI1}
\end{align}
\end{subequations}
For instance, Ref.~\cite{Liu:2019stu} predicts the existence of \textit{twin} (isoscalar) bound states very close (a few $\text{MeV}$) to the $D^{\ast}D^{(\ast)}$ threshold. Based on Eq.~\eqref{eqs:HQSSrel}, for each of the $D^\ast D$ solutions (\textit{isoscalar} or \textit{isovector}) we can predict an additional $D^\ast D^\ast$ state, that we denote $T_{cc}^{\ast}$. If the \textit{isoscalar} solution holds, we predict an $I(J^P)=0(1^+)$ HQSS partner of the $\Tccp$ below the $\DsC\DsN$ threshold, with mass:
\begin{subequations}\label{eq:HQSSIsoscalar}\begin{align}
\Delta M_{{\TccSp}} = M_{{\TccSp}} - m_{\DsC} - m_{\DsN}
& = -1561(71)\ \text{keV}~,\quad [\Lambda=1.0\ \text{GeV}]~,\\
& = -1148(79)\ \text{keV}~,\quad [\Lambda=0.5\ \text{GeV}]~.
\end{align}\end{subequations}
If instead the \textit{isovector} solution is taken, we predict a $1(2^+)$ (isospin triplet), with masses:
\begin{subequations}\label{eq:HQSSIsovector}
\begin{align}
\Delta M_{\TccSpp} = M_{\TccSpp} - 2m_{\DsC}
& = -1580(71)\ \text{keV}~,\quad [\Lambda=1.0\ \text{GeV}]~,\\
\hphantom{\Delta M_{{\TccSp}} = M_{{\TccSp}} - m_{\DsC} - m_{\DsN}}
& = -1156(79)\ \text{keV}~,\quad [\Lambda=0.5\ \text{GeV}]~,
\end{align}
\begin{align}
\Delta M_{{\TccSp}} = M_{{\TccSp}} - m_{\DsC} - m_{\DsN}
& = -1561(71)\ \text{keV}~,\quad [\Lambda=1.0\ \text{GeV}]~,\\
& = -1148(79)\ \text{keV}~,\quad [\Lambda=0.5\ \text{GeV}]~.
\end{align}
\begin{align}
\Delta M_{\TccSz} = M_{\TccSz} - 2m_{\DsN}
& = -1543(71)\ \text{keV}~,\quad [\Lambda=1.0\ \text{GeV}]~,\\
\hphantom{\Delta M_{{\TccSp}} = M_{{\TccSp}} - m_{\DsC} - m_{\DsN}}
& = -1140(79)\ \text{keV}~,\quad [\Lambda=0.5\ \text{GeV}]~.
\end{align}\end{subequations}
Since the decays of the $T_{cc}^\ast$ can be quite involved, we do not predict its width, which should be further investigated. As already explained (Subsec.~\ref{subsec:degeneracy} and ~\ref{subsec:PredictionsSpectra}), the isoscalar solution is favoured, and hence the prediction Eq.~\eqref{eq:HQSSIsoscalar} of an isoscalar state as a HQSS partner of the $\Tccp$ seems the most plausible one.

\section{Conclusions}\label{sec:conclusions}

In this work we have performed a coupled channel analysis of the $D^\ast D$ system, in view of the recent results by the \LHCb collaboration \cite{LHCb:2021vvq,LHCb:2021auc} claiming the existence of a new state, $\Tccp$, seen as a clear signal in the $\DN\DN\pi^+$ spectrum. In our analysis we reproduce the experimental spectrum, and we obtain that the $\Tccp$ appears as a bound state in the $D^\ast D$ amplitude. The bound state shows a large molecular component. While the analysis of Ref.~\cite{LHCb:2021auc} shows no enhancement near the $\DsC\DN$ threshold in the $\DN\DC\pi^+$ spectrum, thus favouring the $I=0$ interpretation of $\Tccp$ (our \textit{isoscalar} solution), more data will be welcome to definitely settle the question. Independently of the nature of the $\Tccp$ state, we have predicted the shape of other $DD\pi$ spectra in which the $\Tccp$ can be seen, and in which the isospin of this state could be determined. Finally, based on Heavy-Quark Spin Symmetry, we have predicted the existence of an isospin singlet or triplet (most likely a singlet, because the \textit{isoscalar} solution seems to be favoured) of $D^\ast D^\ast$ states, $1$--$2\ \text{MeV}$ below the $D^\ast D^\ast$ thresholds.

\section*{Acknowledgements}
We would like to thank E.~Oset for discussions about this work. This work is supported by GVA Grant No. CIDEGENT/2020/002, and by the Spanish Ministerio de Econom\'ia y Competitividad, Ministerio de Ciencia e Innovaci\'on and the European Regional Development Fund (ERDF) under Grants No. PID2019-105439G-C22, No. PID2020-112777GB-I00 (Ref. 10.13039/501100011033), by Generalitat Valenciana under Contract No. PROMETEO/2020/023, and by the EU STRONG-2020 project under the program H2020-INFRAIA2018-1, Grant Agreement No. 824093.

\appendix

\section{\boldmath $K_{t,u}(Q^2)$ functions}\label{app:functions}
For completeness, we show here the specific $K_{t,u}(Q^2)$ of the possible $DD\pi$ final states (with $I=0$ or $1$) discussed in this work, and shown schematically in Figs.~\ref{fig:diagramsDZDZpiP}, \ref{fig:diagramsDPDPpiM}, \ref{fig:diagramsDZDPpiZ}, and \ref{fig:diagramsIzPM}:
\begin{itemize}
\item $\pi^+ D^0 D^0$
\begin{subequations}\begin{align}
K_t(Q^2) & = \alpha \left( 1 + G_{1}(Q^2) T_{11}(Q^2) \right) C_{\DsC \to \DN \pi^+} + \beta G_{2}(Q^2) T_{12}(Q^2)\,  C_{\DsC \to \DN \pi^+}~,\\
K_u(Q^2) & = K_t(Q^2)~.
\end{align}\end{subequations}

\item $\pi^- D^+ D^+$
\begin{subequations}\begin{align}
K_t(Q^2) & = \beta \left( 1 + G_{2}(Q^2) T_{22}(Q^2) \right) C_{\DsN \to \DC \pi^-} + \alpha G_{1}(Q^2) T_{12}(Q^2)\,  C_{\DsN \to \DC \pi^-}~,\\
K_u(Q^2) & = K_t(Q^2)~.
\end{align}\end{subequations}

\item $\pi^0 D^0 D^+$
\begin{subequations}\begin{align}
K_t(Q^2) & = \beta \left( 1 + G_{2}(Q^2) T_{22}(Q^2) \right)  C_{\DsN \to \DN \pi^0} + \alpha G_{1}(Q^2) T_{12}(Q^2)\,  C_{\DsN \to \DN \pi^0}~,\\
K_u(Q^2) & = \alpha \left( 1 + G_{1}(Q^2) T_{11}(Q^2) \right) C_{\DsC \to \DC \pi^0} + \beta G_{2}(Q^2) T_{12}(Q^2)\,  C_{\DsC \to \DC \pi^0}~.
\end{align}\end{subequations}

\item $\pi^0 \DC \DC$
\begin{subequations}\begin{align}
K_t(Q^2) & = \gamma \left( 1 + G_{\DsC\DC}(Q^2) T_{\DsC\DC}(Q^2) \right) C_{\DsC \to \DC \pi^0}~,\\
K_u(Q^2) & = K_t(Q^2)~.
\end{align}\end{subequations}

\item $\pi^+ \DN \DC$
\begin{subequations}\begin{align}
K_t(Q^2) & = \gamma \left( 1 + G_{\DsC\DC}(Q^2) T_{\DsC\DC}(Q^2) \right) C_{\DsC \to \DN \pi^+}~,\\
K_u(Q^2) & = 0~.
\end{align}\end{subequations}

\item $\pi^- \DC \DN$
\begin{subequations}\begin{align}
K_t(Q^2) & = \gamma' \left( 1 + G_{\DsN\DN}(Q^2) T_{\DsN\DN}(Q^2) \right) C_{\DsN \to \DC \pi^-}~,\\
K_u(Q^2) & = 0~.
\end{align}\end{subequations}

\item $\pi^0 \DN \DN$
\begin{subequations}\begin{align}
K_t(Q^2) & = \gamma' \left( 1 + G_{\DsN\DN}(Q^2) T_{\DsN\DN}(Q^2) \right) C_{\DsN \to \DN \pi^0}~.\\
K_u(Q^2) & = K_t(Q^2)~.
\end{align}\end{subequations}

\end{itemize}

\section{Different parameterizations of the interaction kernels}\label{app:vpotGeneralization}
\begin{figure}\centering
\ifthenelse{\boolean{CompileFigures}}{\input{PlotAlternativeFits.tex}}{\includegraphics{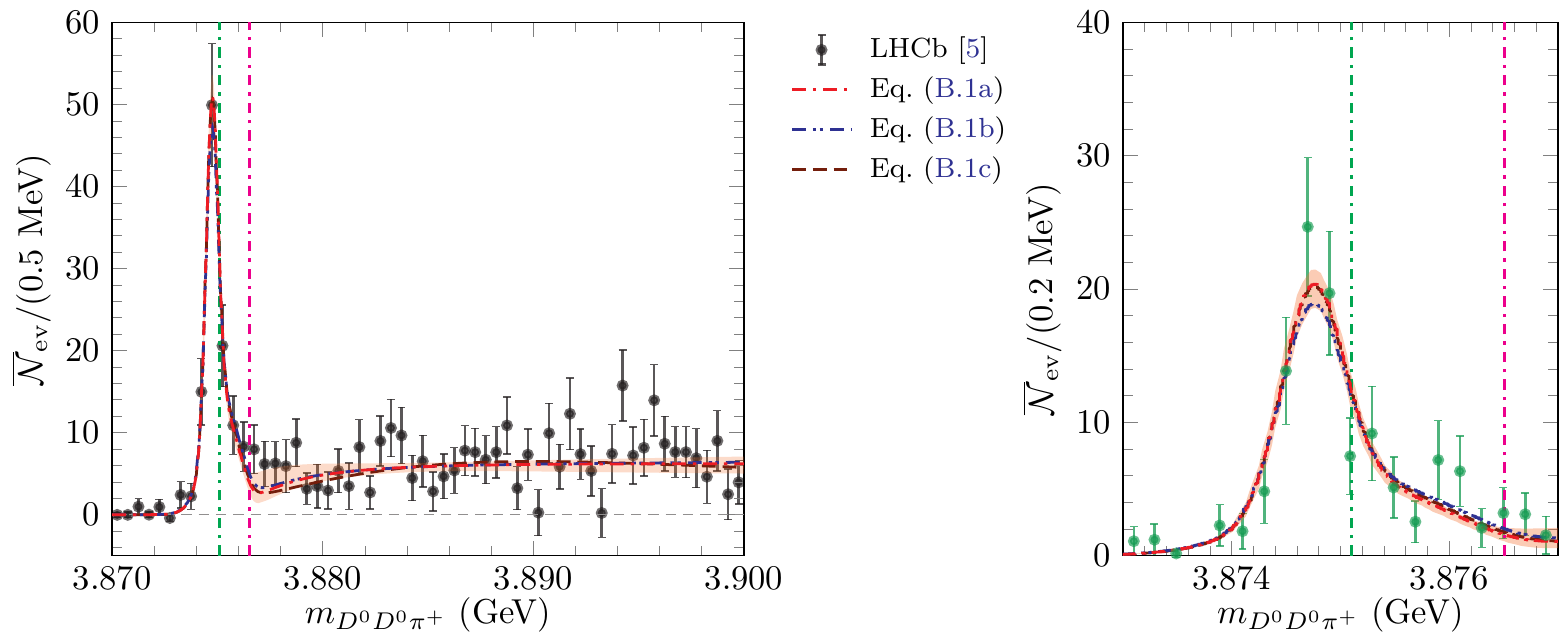}}
\caption{$\DN\DN\pi^+$ spectra obtained for the alternative fits discussed in Appendix~\ref{app:vpotGeneralization}, Eqs.~\eqref{eqs:AltParameterizations}. The pink band represents the uncertainty of the main fit discussed in Subsec.~\ref{subsec:fits}, and is shown here for reference.\label{fig:AlternativeFit}}
\end{figure}

In this Appendix we explore the possibility of using different parameterizations for the kernels $V_{I=0,1}$. For concreteness, we take the case $\Lambda=0.5\ \text{GeV}$ to compare with our main results in Sec.~\ref{sec:results}.\footnote{In the fits that we perform in this Appendix, we restrict ourselves to solutions in which the $\Tccp$ is an isoscalar state (\textit{isoscalar} solutions), keeping in mind that an isovector one can be obtained as discussed in Subsec.~\ref{subsec:degeneracy}.} The variations that we take for our interaction kernels $V_I(s)$ are as follows:
\begin{subequations}\label{eqs:AltParameterizations}\begin{align}
V_0(s) = C_0(\Lambda) & \quad \longrightarrow \quad V_0(s) = C_0(\Lambda) + b_0(\Lambda) k^2~, \label{eq:alt1}\\
V_0(s) = C_0(\Lambda) & \quad \longrightarrow \quad V_0(s) = \frac{d_0(\Lambda)}{E-M_0(\Lambda)}~, \label{eq:alt2}\\
V_1(s) = C_1(\Lambda) & \quad \longrightarrow \quad V_1(s) = C_1(\Lambda) + b_1(\Lambda) k^2~.  \label{eq:alt3}
\end{align}\end{subequations}
The parameters obtained in the fits are:
\begin{itemize}\begin{subequations}\label{eqs:AltFitsParameters}

\item Eq.~\eqref{eq:alt1}:
\begin{align}
& C_0(\Lambda)=-1.52\ \text{fm}^2~,\ b_0(\Lambda)=0.6\ \text{fm}^4~,\ C_1(\Lambda)=-0.54\ \text{fm}^2~,\   \nonumber\\
& \beta/\alpha = 0.098~,\ \chi^2/\text{dof}= 0.94
\end{align}

\item Eq.~\eqref{eq:alt2}:
\begin{align}
& d_0(\Lambda)=-41\ \text{MeV} \cdot \text{fm}^2~,\ M_0 = 3848\ \text{MeV}~,\ C_1(\Lambda)=-0.50\ \text{fm}^2~,\ \nonumber\\
& \beta/\alpha = 0.160~,\ \chi^2/\text{dof}= 0.90
\end{align}

\item Eq.~\eqref{eq:alt3}:
\begin{align}
& C_0(\Lambda)=-1.55\ \text{fm}^2~,\ C_1(\Lambda)=-0.54\ \text{fm}^2~,\ b_1(\Lambda)=-0.7\ \text{fm}^4~,\   \nonumber\\
& \beta/\alpha = 0.030~,\ \chi^2/\text{dof}= 0.94
\end{align}
\end{subequations}\end{itemize}
The resulting spectrum for each of these fits is shown in Fig.~\ref{fig:AlternativeFit}, and as can be seen they all lie essentially within the uncertainty band of the main fit discussed in Subsec.~\ref{subsec:fits}, which justifies taking our fit in Subsec.~\ref{subsec:fits} as our main one. Some of the output quantities ($\Tccp$ mass and width, etc.) are shown in Table~\ref{tab:altfits} for the different fits. The fitted parameters in Eqs.~\eqref{eqs:AltFitsParameters} can show deviations across the fits with respect to the main fit in Table~\ref{table:parameters} which are larger than the statistical uncertainty. \textit{E.g.}, $C_0(\Lambda)=-1.5417(121)\ \text{fm}^2$ in the main fit, but here one can have deviations of about $0.03\ \text{fm}^2$, larger than the statistical one, which is $\sim 0.01\ \text{fm}^2$. This seems to be the case too for the width and the couplings of the $\Tccp$, but not for the mass. If we now pay attention to the molecular probability, calculated in the isospin limit as Eq.~\eqref{eq:PIWeinIsLim} (last row in Table~\ref{tab:altfits}), we see that in all cases we still have $P_I \geqslant 0.75$, \textit{i.e.} even when more complex parameterizations are allowed one still finds a very large molecular component.

\begin{table}\centering
\begin{tabular}{ccccc} \hline
 & \eqref{eq:alt1} & \eqref{eq:alt2} & \eqref{eq:alt3} & Main \\ \hline\hline
$\Delta M_{\Tccp}\ \text{(keV)}$ & $-350$ & $-357$ & $-355$ & $-356(29)$ \\
$\Gamma\ \text{(keV)}$ & $78$ & $106$ & $77$ & $78(1)$ \\ 
$g_{\cho}\ \text{(GeV)}$ & $4.61$   &   $5.03$         &  $4.27$ & $4.36(16)$ \\
$g_{\cht}\ \text{(GeV)}$ & $4.24$   &   $4.68$         &  $3.97$ & $3.67(50)$ \\
$a_{\cho}\ \text{(fm)}$  & $-9.26+2.98i$ & $-9.87+3.59i$ & $-8.36+2.48i$ & $-8.56(49)+2.61(32)i$ \\  \hline
$P_{\cho}$ & $0.89$  &   $1.05$        &  $0.76$         & $0.79(5)$ \\
$P_{\cht}$ & $0.28$  &   $0.34$        &  $0.24$         & $0.21(5)$ \\ \hline
$a_{I}\ \text{(fm)}$     & $-6.27$     & $-6.65$ & $-5.29$ & $-5.57(25)$ \\
$r_{I}\ \text{(fm)}$     & $1.95$      & $2.51$  & $1.26$  & $1.26$ \\
$P_{I}$ [Eq.~\eqref{eq:PIWeinIsLim}]  & $0.79$ & $0.76$ & $0.82$ & $0.830(5)$ \\ \hline
\end{tabular}
\caption{Values of different output quantities discussed in Sec.~\ref{sec:results} when the different parameterizations Eqs.~\eqref{eqs:AltParameterizations} are employed. In the last column, for reference, we show the result obtained in the main fit.\label{tab:altfits}}
\end{table}

\bibliographystyle{JHEP}
\bibliography{TccPlus}

\end{document}

%% file: TikZsettings.tex
\pgfdeclarepatternformonly{crosshatchD}{\pgfqpoint{-1.5pt}{-1.5pt}}{\pgfqpoint{4pt}{4pt}}{\pgfqpoint{2.5pt}{2.5pt}}%
{%
  \pgfsetlinewidth{0.4pt}%
  \pgfpathmoveto{\pgfqpoint{3.1pt}{0pt}}%
  \pgfpathlineto{\pgfqpoint{0pt}{3.1pt}}%
  \pgfpathmoveto{\pgfqpoint{0pt}{0pt}}%
  \pgfpathlineto{\pgfqpoint{3.1pt}{3.1pt}}%
  \pgfusepath{stroke}%
}%

\tikzset{vpiFF/.style={draw=Red,fill=Pink!40 ,circle,very thick,inner sep=4.5pt}}
\tikzset{vfone/.style={draw=Blue,fill=Aqua!40,circle,very thick,inner sep=4.5pt}}

\tikzset{VertexSource/.style={draw=Blue,fill=Blue!40 ,circle,crossed dot,very thick,inner sep=2pt}}
\tikzset{VDsDpi/.style={draw=Green,fill=Green,circle,very thick,inner sep=2pt}}
\tikzset{VertexTMat/.style={draw=Red,fill=Red!40 ,rectangle,very thick,inner sep=3pt}}

\tikzfeynmanset{DVec/.style={Black,double,very thick}}
\tikzfeynmanset{DMes/.style={Black,very thick}}
\tikzfeynmanset{pion/.style={Black,dash pattern=on 3pt off 2pt,very thick}}

\tikzfeynmanset{phot/.style={black,photon,very thick}}
\tikzfeynmanset{omeg/.style={black,fermion,very thick}}
\tikzfeynmanset{cutAs/.style={black!80,very thick}}
\tikzfeynmanset{cutBs/.style={black!80,thin,decoration={pre length=0.06cm, post length=0.0cm,border,segment length=0.4mm,amplitude=1mm,angle=270},
            postaction={decorate,draw}}}

\pgfplotsset{error bar legend/.style={%
    /pgfplots/legend image code/.prefix code={%
      \pgfkeysgetvalue{/pgfplots/error bars/error mark}{\pgfplotserrorbarsmark}%
      \draw[%
        /pgfplots/every error bar, 
        mark=\pgfplotserrorbarsmark, 
        /pgfplots/error bars/error mark options, 
        sharp plot,
        ##1,
      ] plot coordinates {(0.30cm, -0.15cm) (0.30cm, 0.15cm)};%
    }
  }
}

\pgfplotsset{MyFit/.style={thick,no markers, Blue}}
\pgfplotsset{MyFitAB/.style={thick,no markers, Orange}}
\pgfplotsset{AlternativeFit/.style={thick,no markers}}

\pgfplotsset{AlternativeFitOne/.style={AlternativeFit, Red ,dash pattern=on 4pt off 2pt on 1pt off 2pt}}
\pgfplotsset{AlternativeFitTwo/.style={AlternativeFit, Blue,dash pattern=on 4pt off 2pt on 1pt off 1pt on 1pt off 2pt}}
\pgfplotsset{AlternativeFitThr/.style={AlternativeFit, Brown,dash pattern=on 4pt off 2pt}}

\tikzset{FrstThr/.style={draw,thick,Green  ,dash pattern=on 3pt off 2pt on 1pt off 2pt}}
\tikzset{ScndThr/.style={draw,thick,Magenta,dash pattern=on 3pt off 2pt on 1pt off 2pt}}
\tikzset{Zero/.style={draw,Gray,dash pattern=on 3pt off 2pt}}

\pgfplotsset{SpIzZeroA/.style={very thick,no markers, Blue}}
\pgfplotsset{SpIzZeroB/.style={very thick,no markers, Orange, dash pattern=on 3pt off 2pt on 1pt off 2pt}}
\pgfplotsset{SpIzZeroC/.style={very thick,no markers, Red, dash pattern=on 4pt off 2pt}}

\pgfplotsset{IsoscalarSol/.style={very thick,no markers,opacity=0.8}}
\pgfplotsset{IsovectorSol/.style={thick,no markers,dash pattern=on 3.5pt off 2.25pt,opacity=0.8}}

\pgfplotsset{SpIzOneAIsovectorSol/.style={Blue,IsovectorSol}}
\pgfplotsset{SpIzOneBIsovectorSol/.style={Brown,IsovectorSol}}
\pgfplotsset{SpIzOneCIsovectorSol/.style={Red,IsovectorSol}}
\pgfplotsset{SpIzOneAIsoscalarSol/.style={Blue,IsoscalarSol}}
\pgfplotsset{SpIzOneBIsoscalarSol/.style={Brown,IsoscalarSol}}\pgfplotsset{SpIzOneCIsoscalarSol/.style={Red,IsoscalarSol}}

\pgfplotsset{ReDZO/.style={very thick,no markers}}
\pgfplotsset{ImDZO/.style={very thick,no markers,dash pattern=on 3.5pt off 1.75pt on 1pt off 1.75pt}}
\pgfplotsset{ReDZ/.style={ReDZO,Blue}}
\pgfplotsset{ImDZ/.style={ImDZO,Red}}
\pgfplotsset{ReDO/.style={ReDZO,Green}}
\pgfplotsset{ImDO/.style={ImDZO,Orange}}

\pgfplotsset{DataLHCb/.style={line width=1.1pt,color=Black,only marks, mark=*,mark size=1.15pt,draw opacity=0.65, fill opacity=0.75,on layer={axis background},error bar legend,error bars/.cd, y dir=both,y explicit,error bar style={draw opacity=0.75,fill opacity=0.75,line width=0.75pt}}}

\pgfplotsset{DataLHCbInset/.style={line width=1.1pt,color=ForestGreen,only marks, mark=*,mark size=1.15pt,draw opacity=0.65, fill opacity=0.75,on layer={axis background},error bar legend,error bars/.cd, y dir=both,y explicit,error bar style={draw opacity=0.75,fill opacity=0.75,line width=0.75pt}}}

\pgfplotsset{PullLar/.style={line width=1.1pt,color=Black,only marks, mark=*,mark size=0.85pt,draw opacity=0.65, fill opacity=0.75,on layer={axis background},error bar legend,error bars/.cd, y dir=both,y explicit,error bar style={draw opacity=0.75,fill opacity=0.75,line width=0.75pt}}}

\pgfplotsset{PullSma/.style={line width=1.1pt,color=ForestGreen,only marks, mark=*,mark size=0.85pt,draw opacity=0.65, fill opacity=0.75,on layer={axis background},error bar legend,error bars/.cd, y dir=both,y explicit,error bar style={draw opacity=0.75,fill opacity=0.75,line width=0.75pt}}}

\pgfplotsset{PullLarMC/.style={line width=1pt,color=Purple,only marks, mark=square,mark size=0.75pt,draw opacity=0.65, fill opacity=0.75,on layer={axis background},error bar legend,error bars/.cd, y dir=both,y explicit,error bar style={draw opacity=0.75,fill opacity=0.75,line width=0.75pt}}}

\pgfplotsset{PullSmaMC/.style={line width=1pt,color=Orange,only marks, mark=square,mark size=0.75pt,draw opacity=0.65, fill opacity=0.75,on layer={axis background},error bar legend,error bars/.cd, y dir=both,y explicit,error bar style={draw opacity=0.75,fill opacity=0.75,line width=0.75pt}}}

\tikzset{declare function={ 
G(\x,\av,\sd) = 1.0/sqrt(2*pi*\sd^2) * exp(-(\x-\av)^2/(2*\sd^2));
}}

\pgfplotsset{HistoStyleL/.style={const plot mark mid,draw=none,fill=lowphi,opacity=0.5}}
\pgfplotsset{HistoStyleH/.style={const plot mark mid,draw=none,fill=higphi,opacity=0.5}}

\pgfmathsetmacro{\onesigmatwopar}{1.515}

\tikzset{
        vertical align/.style={
            baseline=-.5*(height("$+$")-depth("$+$"))
        }}        
\newcommand{\DDpiDiagramTree}[5]{%
\begin{tikzpicture}[vertical align]\footnotesize
\begin{feynman}
\vertex [VertexSource,label={left:#1}] (VS) {};
\vertex [VDsDpi,above right=20pt and 30pt of VS] (TreeVDsDpi) {};
\vertex [         below right=20pt and 30pt of VS,label={right:#5}] (DOutBot) {};
\vertex [above right=20pt and 20pt of TreeVDsDpi,label={right:#4}] (DOutTop) {};
\vertex [below right=20pt and 20pt of TreeVDsDpi,label={right:#3}] (piOut) {};

\diagram* { 
(VS) -- [DVec,edge label'=#2] (TreeVDsDpi),%
(VS) -- [DMes] (DOutBot),%
(TreeVDsDpi) -- [pion] (piOut),%
(TreeVDsDpi) -- [DMes] (DOutTop),%
};
\end{feynman}
\end{tikzpicture}%
}

\newcommand{\DDpiDiagramLoop}[7]{%
\begin{tikzpicture}[vertical align]\footnotesize
\begin{feynman}
\vertex [VertexSource,label={left:#1}] (VS) {};
\vertex [VertexTMat, right=30pt of VS] (TMat) {};
\vertex [VDsDpi,above right=20pt and 30pt of TMat] (TreeVDsDpi) {};
\node [         below right=20pt and 30pt of TMat,label={right:#5}] (DOutBot) {};
\node [above right=20pt and 20pt of TreeVDsDpi,label={right:#4}] (DOutTop) {};
\node [below right=20pt and 20pt of TreeVDsDpi,label={right:#3}] (piOut) {};

\diagram* {
(VS) -- [DVec,out=+45,in=+135,edge label =#6]  (TMat),%
(VS) -- [DMes,out=-45,in=-135,edge label'=#7] (TMat),%
(TMat) -- [DVec,edge label'=#2]  (TreeVDsDpi),%
(TMat) -- [DMes] (DOutBot),%
(TreeVDsDpi) -- [pion] (piOut),%
(TreeVDsDpi) -- [DMes] (DOutTop),%
};
\end{feynman}
\end{tikzpicture}%
}

%% file: shorthands.tex
\newcommand{\braket}[3]{\left\langle #1 \right\rvert #2 \left\lvert #3 \right\rangle}

\newcommand\bsub{\begin{subequations}}
\newcommand\esub{\end{subequations}}

\def\XXint#1#2#3{{\setbox0=\hbox{$#1{#2#3}{\int}$}\vcenter{\hbox{$#2#3$}}\kern-.5\wd0}}

%% file: DiagramsOtherIsovector.tex
\begin{tabular}{|c|c|} \hline
\ifthenelse{\boolean{CompileDiagrams}}{\tikzsetnextfilename{DiagramsEa1}
\DDpiDiagramTree{$\gamma$}{$\DsC$}{$\pi^0$}{$\DC$}{$\DC$}}{\raisebox{-0.33\height}{\includegraphics{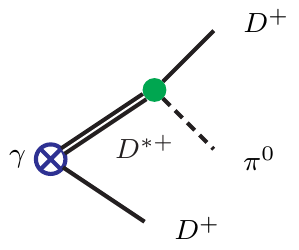}}}
$+$%
\ifthenelse{\boolean{CompileDiagrams}}{\tikzsetnextfilename{DiagramsEa2}
\DDpiDiagramLoop{$\gamma$}{$\DsC$}{$\pi^0$}{$\DC$}{$\DC$}{$\DsC$}{$\DC$}}{\raisebox{-0.33\height}{\includegraphics{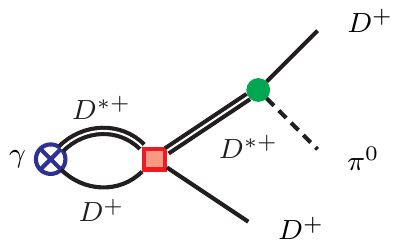}}}%
& %
\ifthenelse{\boolean{CompileDiagrams}}{\tikzsetnextfilename{DiagramsFa1}
\DDpiDiagramTree{$\gamma'$}{$\DsN$}{$\pi^-$}{$\DC$}{$\DN$}}{\raisebox{-0.33\height}{\includegraphics{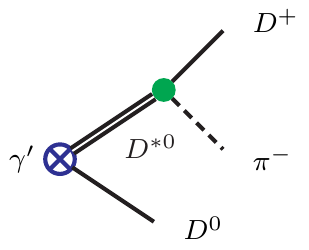}}}
$+$%
\ifthenelse{\boolean{CompileDiagrams}}{\tikzsetnextfilename{DiagramsFa2}
\DDpiDiagramLoop{$\gamma'$}{$\DsN$}{$\pi^-$}{$\DC$}{$\DN$}{$\DsN$}{$\DN$}}{\raisebox{-0.33\height}{\includegraphics{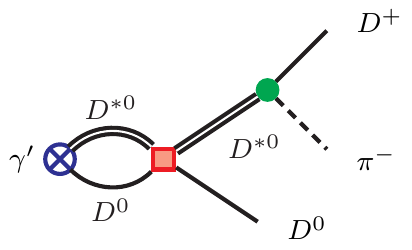}}}%
\\ \hline
\ifthenelse{\boolean{CompileDiagrams}}{\tikzsetnextfilename{DiagramsEb1}
\DDpiDiagramTree{$\gamma$}{$\DsC$}{$\pi^+$}{$\DN$}{$\DC$}}{\raisebox{-0.33\height}{\includegraphics{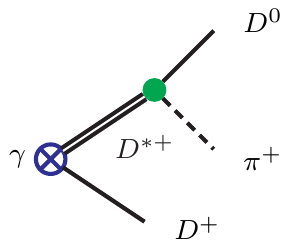}}}
$+$%
\ifthenelse{\boolean{CompileDiagrams}}{\tikzsetnextfilename{DiagramsEb2}
\DDpiDiagramLoop{$\gamma$}{$\DsC$}{$\pi^+$}{$\DN$}{$\DC$}{$\DsC$}{$\DC$}}{\raisebox{-0.33\height}{\includegraphics{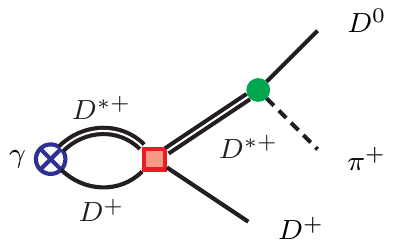}}}%
 & %
\ifthenelse{\boolean{CompileDiagrams}}{\tikzsetnextfilename{DiagramsFb1}
\DDpiDiagramTree{$\gamma'$}{$\DsN$}{$\pi^0$}{$\DN$}{$\DN$}}{\raisebox{-0.33\height}{\includegraphics{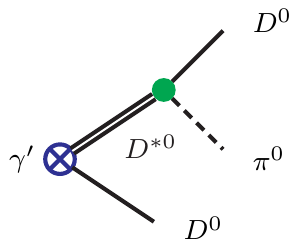}}}
$+$%
\ifthenelse{\boolean{CompileDiagrams}}{\tikzsetnextfilename{DiagramsFb2}
\DDpiDiagramLoop{$\gamma'$}{$\DsN$}{$\pi^0$}{$\DN$}{$\DN$}{$\DsN$}{$\DN$}}{\raisebox{-0.33\height}{\includegraphics{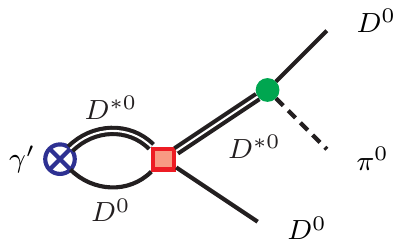}}}%
\\ \hline
 \end{tabular}

%% file: main.bbl
\providecommand{\href}[2]{#2}\begingroup\raggedright\begin{thebibliography}{10}

\bibitem{Gell-Mann:1964ewy}
M.~Gell-Mann, \emph{{A Schematic Model of Baryons and Mesons}},
  \href{https://doi.org/10.1016/S0031-9163(64)92001-3}{\emph{Phys. Lett.}
  {\bfseries 8} (1964) 214}.

\bibitem{Zweig:1964ruk}
G.~Zweig, \emph{{An SU(3) model for strong interaction symmetry and its
  breaking. Version 1}},  CERN-TH-401.

\bibitem{Zweig:1964jf}
G.~Zweig, \emph{{An SU(3) model for strong interaction symmetry and its
  breaking. Version 2}},  in \emph{{DEVELOPMENTS IN THE QUARK THEORY OF
  HADRONS. VOL. 1. 1964 - 1978}}, D.B.~Lichtenberg and S.P.~Rosen, eds.,
  pp.~22--101 (1964).

\bibitem{Godfrey:1985xj}
S.~Godfrey and N.~Isgur, \emph{{Mesons in a Relativized Quark Model with
  Chromodynamics}}, \href{https://doi.org/10.1103/PhysRevD.32.189}{\emph{Phys.
  Rev. D} {\bfseries 32} (1985) 189}.

\bibitem{LHCb:2021vvq}
R.~Aaij et~al., {\scshape LHCb} collaboration, \emph{{Observation of an exotic
  narrow doubly charmed tetraquark}},
  \href{https://arxiv.org/abs/2109.01038}{{\ttfamily 2109.01038}}.

\bibitem{LHCb:2021auc}
R.~Aaij et~al., {\scshape LHCb} collaboration, \emph{{Study of the doubly
  charmed tetraquark $T_{cc}^+$}},
  \href{https://arxiv.org/abs/2109.01056}{{\ttfamily 2109.01056}}.

\bibitem{LHCb:2017iph}
R.~Aaij et~al., {\scshape LHCb} collaboration, \emph{{Observation of the doubly
  charmed baryon $\Xi_{cc}^{++}$}},
  \href{https://doi.org/10.1103/PhysRevLett.119.112001}{\emph{Phys. Rev. Lett.}
  {\bfseries 119} (2017) 112001}
  [\href{https://arxiv.org/abs/1707.01621}{{\ttfamily 1707.01621}}].

\bibitem{Dong:2021bvy}
X.-K.~Dong, F.-K.~Guo and B.-S.~Zou, \emph{{A survey of heavy-heavy hadronic
  molecules}},  \href{https://arxiv.org/abs/2108.02673}{{\ttfamily
  2108.02673}}.

\bibitem{Li:2021zbw}
N.~Li, Z.-F.~Sun, X.~Liu and S.-L.~Zhu, \emph{{Perfect $DD^*$ molecular
  prediction matching the $T_{cc}$ observation at LHCb}},
  \href{https://doi.org/10.1088/0256-307X/38/9/092001}{\emph{Chin. Phys. Lett.}
  {\bfseries 38} (2021) 092001}
  [\href{https://arxiv.org/abs/2107.13748}{{\ttfamily 2107.13748}}].

\bibitem{Meng:2021jnw}
L.~Meng, G.-J.~Wang, B.~Wang and S.-L.~Zhu, \emph{{Probing the long-range
  structure of the Tcc+ with the strong and electromagnetic decays}},
  \href{https://doi.org/10.1103/PhysRevD.104.L051502}{\emph{Phys. Rev. D}
  {\bfseries 104} (2021) 051502}
  [\href{https://arxiv.org/abs/2107.14784}{{\ttfamily 2107.14784}}].

\bibitem{Dai:2021wxi}
L.-Y.~Dai, X.~Sun, X.-W.~Kang, A.P.~Szczepaniak and J.-S.~Yu, \emph{{Pole
  analysis on the doubly charmed meson in $D^0D^0\pi^+$ mass spectrum}},
  \href{https://arxiv.org/abs/2108.06002}{{\ttfamily 2108.06002}}.

\bibitem{Feijoo:2021ppq}
A.~Feijoo, W.H.~Liang and E.~Oset, \emph{{$D^0 D^0 \pi^+$ mass distribution in
  the production of the $T_{cc}$ exotic state}},
  \href{https://arxiv.org/abs/2108.02730}{{\ttfamily 2108.02730}}.

\bibitem{Agaev:2021vur}
S.S.~Agaev, K.~Azizi and H.~Sundu, \emph{{Newly observed exotic doubly charmed
  meson $T^{+}_{cc}$}},  \href{https://arxiv.org/abs/2108.00188}{{\ttfamily
  2108.00188}}.

\bibitem{Wu:2021kbu}
T.-W.~Wu, Y.-W.~Pan, M.-Z.~Liu, S.-Q.~Luo, X.~Liu and L.-S.~Geng,
  \emph{{Discovery of the doubly charmed $T_{cc}^+$ state implies a triply
  charmed $H_{ccc}$ hexaquark state}},
  \href{https://arxiv.org/abs/2108.00923}{{\ttfamily 2108.00923}}.

\bibitem{Ling:2021bir}
X.-Z.~Ling, M.-Z.~Liu, L.-S.~Geng, E.~Wang and J.-J.~Xie, \emph{{Can we
  understand the decay width of the $T_{cc}^+$ state?}},
  \href{https://arxiv.org/abs/2108.00947}{{\ttfamily 2108.00947}}.

\bibitem{Chen:2021vhg}
R.~Chen, Q.~Huang, X.~Liu and S.-L.~Zhu, \emph{{Another doubly charmed
  molecular resonance $T_{cc}^{\prime+}(3876)$}},
  \href{https://arxiv.org/abs/2108.01911}{{\ttfamily 2108.01911}}.

\bibitem{Yan:2021wdl}
M.-J.~Yan and M.P.~Valderrama, \emph{{Subleading contributions to the decay
  width of the $T_{cc}^+$ tetraquark}},
  \href{https://arxiv.org/abs/2108.04785}{{\ttfamily 2108.04785}}.

\bibitem{Weng:2021hje}
X.-Z.~Weng, W.-Z.~Deng and S.-L.~Zhu, \emph{{Doubly heavy tetraquarks in an
  extended chromomagnetic model}},
  \href{https://arxiv.org/abs/2108.07242}{{\ttfamily 2108.07242}}.

\bibitem{Huang:2021urd}
Y.~Huang, H.Q.~Zhu, L.-S.~Geng and R.~Wang, \emph{{Production of the
  $T^{+}_{cc}$ state in the $\gamma{}p\to{}D^{+}\bar{T}^{-}_{cc}\Lambda_c^{+}$
  reaction}},  \href{https://arxiv.org/abs/2108.13028}{{\ttfamily 2108.13028}}.

\bibitem{Chen:2021kad}
R.~Chen, N.~Li, Z.-F.~Sun, X.~Liu and S.-L.~Zhu, \emph{{Doubly charmed
  molecular pentaquarks}},
  \href{https://doi.org/10.1016/j.physletb.2021.136693}{\emph{Phys. Lett. B}
  {\bfseries 822} (2021) 136693}
  [\href{https://arxiv.org/abs/2108.12730}{{\ttfamily 2108.12730}}].

\bibitem{Xin:2021wcr}
Q.~Xin and Z.-G.~Wang, \emph{{Analysis of the axialvector doubly-charmed
  tetraquark molecular states with the QCD sum rules}},
  \href{https://arxiv.org/abs/2108.12597}{{\ttfamily 2108.12597}}.

\bibitem{Fleming:2021wmk}
S.~Fleming, R.~Hodges and T.~Mehen, \emph{{$T_{cc}^+$ decays: differential
  spectra and two-body final states}},
  \href{https://arxiv.org/abs/2109.02188}{{\ttfamily 2109.02188}}.

\bibitem{Chen:2021tnn}
X.~Chen, \emph{{Doubly heavy tetraquark states $cc\bar{u}\bar{d}$ and
  $bb\bar{u}\bar{d}$}},  \href{https://arxiv.org/abs/2109.02828}{{\ttfamily
  2109.02828}}.

\bibitem{Azizi:2021aib}
K.~Azizi and U.~\"Ozdem, \emph{{Magnetic dipole moments of the $T_{cc}^+$ and
  $Z_V^{++}$ tetraquark states}},
  \href{https://arxiv.org/abs/2109.02390}{{\ttfamily 2109.02390}}.

\bibitem{Ren:2021dsi}
H.~Ren, F.~Wu and R.~Zhu, \emph{{Hadronic molecule interpretation of $T^+_{cc}$
  and its beauty-partners}},
  \href{https://arxiv.org/abs/2109.02531}{{\ttfamily 2109.02531}}.

\bibitem{Yang:2021zhe}
G.~Yang, J.~Ping and J.~Segovia, \emph{{Hidden-charm tetraquarks with
  strangeness in the chiral quark model}},
  \href{https://arxiv.org/abs/2109.04311}{{\ttfamily 2109.04311}}.

\bibitem{Jin:2021cxj}
Y.~Jin, S.-Y.~Li, Y.-R.~Liu, Q.~Qin, Z.-G.~Si and F.-S.~Yu, \emph{{Colour and
  baryon number fluctuation of preconfinement system in production process and
  $T_{cc}$ structure}},  \href{https://arxiv.org/abs/2109.05678}{{\ttfamily
  2109.05678}}.

\bibitem{Hu:2021gdg}
Y.~Hu, J.~Liao, E.~Wang, Q.~Wang, H.~Xing and H.~Zhang, \emph{{The production
  of doubly charmed exotic hadrons in heavy ion collisions}},
  \href{https://arxiv.org/abs/2109.07733}{{\ttfamily 2109.07733}}.

\bibitem{Chen:2021cfl}
K.~Chen, R.~Chen, L.~Meng, B.~Wang and S.-L.~Zhu, \emph{{Systematics of the
  heavy flavor hadronic molecules}},
  \href{https://arxiv.org/abs/2109.13057}{{\ttfamily 2109.13057}}.

\bibitem{He:2021smz}
J.~He, D.-Y.~Chen, Z.-W.~Liu and X.~Liu, \emph{{New clean fission with hadronic
  molecular states}},  \href{https://arxiv.org/abs/2109.14395}{{\ttfamily
  2109.14395}}.

\bibitem{Junnarkar:2018twb}
P.~Junnarkar, N.~Mathur and M.~Padmanath, \emph{{Study of doubly heavy
  tetraquarks in Lattice QCD}},
  \href{https://doi.org/10.1103/PhysRevD.99.034507}{\emph{Phys. Rev. D}
  {\bfseries 99} (2019) 034507}
  [\href{https://arxiv.org/abs/1810.12285}{{\ttfamily 1810.12285}}].

\bibitem{Ikeda:2013vwa}
Y.~Ikeda, B.~Charron, S.~Aoki, T.~Doi, T.~Hatsuda, T.~Inoue et~al.,
  \emph{{Charmed tetraquarks $T_{cc}$ and $T_{cs}$ from dynamical lattice QCD
  simulations}},
  \href{https://doi.org/10.1016/j.physletb.2014.01.002}{\emph{Phys. Lett. B}
  {\bfseries 729} (2014) 85} [\href{https://arxiv.org/abs/1311.6214}{{\ttfamily
  1311.6214}}].

\bibitem{Cheung:2017tnt}
G.K.C.~Cheung, C.E.~Thomas, J.J.~Dudek and R.G.~Edwards, {\scshape Hadron
  Spectrum} collaboration, \emph{{Tetraquark operators in lattice QCD and
  exotic flavour states in the charm sector}},
  \href{https://doi.org/10.1007/JHEP11(2017)033}{\emph{JHEP} {\bfseries 11}
  (2017) 033} [\href{https://arxiv.org/abs/1709.01417}{{\ttfamily
  1709.01417}}].

\bibitem{Francis:2018jyb}
A.~Francis, R.J.~Hudspith, R.~Lewis and K.~Maltman, \emph{{Evidence for
  charm-bottom tetraquarks and the mass dependence of heavy-light tetraquark
  states from lattice QCD}},
  \href{https://doi.org/10.1103/PhysRevD.99.054505}{\emph{Phys. Rev. D}
  {\bfseries 99} (2019) 054505}
  [\href{https://arxiv.org/abs/1810.10550}{{\ttfamily 1810.10550}}].

\bibitem{Leskovec:2019ioa}
L.~Leskovec, S.~Meinel, M.~Pflaumer and M.~Wagner, \emph{{Lattice QCD
  investigation of a doubly-bottom $\bar{b} \bar{b} u d$ tetraquark with
  quantum numbers $I(J^P) = 0(1^+)$}},
  \href{https://doi.org/10.1103/PhysRevD.100.014503}{\emph{Phys. Rev. D}
  {\bfseries 100} (2019) 014503}
  [\href{https://arxiv.org/abs/1904.04197}{{\ttfamily 1904.04197}}].

\bibitem{ParticleDataGroup:2020ssz}
P.A.~Zyla et~al., {\scshape Particle Data Group} collaboration, \emph{{Review
  of Particle Physics}},
  \href{https://doi.org/10.1093/ptep/ptaa104}{\emph{PTEP} {\bfseries 2020}
  (2020) 083C01}.

\bibitem{James:1994vla}
F.~James, \emph{{MINUIT Function Minimization and Error Analysis: Reference
  Manual Version 94.1}}, .

\bibitem{Weinberg:1965zz}
S.~Weinberg, \emph{{Evidence That the Deuteron Is Not an Elementary Particle}},
  \href{https://doi.org/10.1103/PhysRev.137.B672}{\emph{Phys. Rev.} {\bfseries
  137} (1965) B672}.

\bibitem{Baru:2003qq}
V.~Baru, J.~Haidenbauer, C.~Hanhart, Y.~Kalashnikova and A.E.~Kudryavtsev,
  \emph{{Evidence that the a(0)(980) and f(0)(980) are not elementary
  particles}},
  \href{https://doi.org/10.1016/j.physletb.2004.01.088}{\emph{Phys. Lett. B}
  {\bfseries 586} (2004) 53}
  [\href{https://arxiv.org/abs/hep-ph/0308129}{{\ttfamily hep-ph/0308129}}].

\bibitem{Gamermann:2009uq}
D.~Gamermann, J.~Nieves, E.~Oset and E.~Ruiz~Arriola, \emph{{Couplings in
  coupled channels versus wave functions: application to the X(3872)
  resonance}}, \href{https://doi.org/10.1103/PhysRevD.81.014029}{\emph{Phys.
  Rev. D} {\bfseries 81} (2010) 014029}
  [\href{https://arxiv.org/abs/0911.4407}{{\ttfamily 0911.4407}}].

\bibitem{Yamagata-Sekihara:2010kpd}
J.~Yamagata-Sekihara, J.~Nieves and E.~Oset, \emph{{Couplings in coupled
  channels versus wave functions in the case of resonances: application to the
  two $\Lambda(1405)$ states}},
  \href{https://doi.org/10.1103/PhysRevD.83.014003}{\emph{Phys. Rev. D}
  {\bfseries 83} (2011) 014003}
  [\href{https://arxiv.org/abs/1007.3923}{{\ttfamily 1007.3923}}].

\bibitem{Hyodo:2011qc}
T.~Hyodo, D.~Jido and A.~Hosaka, \emph{{Compositeness of dynamically generated
  states in a chiral unitary approach}},
  \href{https://doi.org/10.1103/PhysRevC.85.015201}{\emph{Phys. Rev. C}
  {\bfseries 85} (2012) 015201}
  [\href{https://arxiv.org/abs/1108.5524}{{\ttfamily 1108.5524}}].

\bibitem{Aceti:2012dd}
F.~Aceti and E.~Oset, \emph{{Wave functions of composite hadron states and
  relationship to couplings of scattering amplitudes for general partial
  waves}}, \href{https://doi.org/10.1103/PhysRevD.86.014012}{\emph{Phys. Rev.
  D} {\bfseries 86} (2012) 014012}
  [\href{https://arxiv.org/abs/1202.4607}{{\ttfamily 1202.4607}}].

\bibitem{Guo:2015daa}
Z.-H.~Guo and J.A.~Oller, \emph{{Probabilistic interpretation of compositeness
  relation for resonances}},
  \href{https://doi.org/10.1103/PhysRevD.93.096001}{\emph{Phys. Rev. D}
  {\bfseries 93} (2016) 096001}
  [\href{https://arxiv.org/abs/1508.06400}{{\ttfamily 1508.06400}}].

\bibitem{Sekihara:2014kya}
T.~Sekihara, T.~Hyodo and D.~Jido, \emph{{Comprehensive analysis of the wave
  function of a hadronic resonance and its compositeness}},
  \href{https://doi.org/10.1093/ptep/ptv081}{\emph{PTEP} {\bfseries 2015}
  (2015) 063D04} [\href{https://arxiv.org/abs/1411.2308}{{\ttfamily
  1411.2308}}].

\bibitem{Oller:2017alp}
J.A.~Oller, \emph{{New results from a number operator interpretation of the
  compositeness of bound and resonant states}},
  \href{https://doi.org/10.1016/j.aop.2018.07.023}{\emph{Annals Phys.}
  {\bfseries 396} (2018) 429}
  [\href{https://arxiv.org/abs/1710.00991}{{\ttfamily 1710.00991}}].

\bibitem{Matuschek:2020gqe}
I.~Matuschek, V.~Baru, F.-K.~Guo and C.~Hanhart, \emph{{On the nature of
  near-threshold bound and virtual states}},
  \href{https://doi.org/10.1140/epja/s10050-021-00413-y}{\emph{Eur. Phys. J. A}
  {\bfseries 57} (2021) 101}
  [\href{https://arxiv.org/abs/2007.05329}{{\ttfamily 2007.05329}}].

\bibitem{Neubert:1993mb}
M.~Neubert, \emph{{Heavy quark symmetry}},
  \href{https://doi.org/10.1016/0370-1573(94)90091-4}{\emph{Phys. Rept.}
  {\bfseries 245} (1994) 259}
  [\href{https://arxiv.org/abs/hep-ph/9306320}{{\ttfamily hep-ph/9306320}}].

\bibitem{Manohar:2000dt}
A.V.~Manohar and M.B.~Wise, \emph{{Heavy quark physics}}, vol.~10 (2000).

\bibitem{Hidalgo-Duque:2012rqv}
C.~Hidalgo-Duque, J.~Nieves and M.P.~Valderrama, \emph{{Light flavor and heavy
  quark spin symmetry in heavy meson molecules}},
  \href{https://doi.org/10.1103/PhysRevD.87.076006}{\emph{Phys. Rev. D}
  {\bfseries 87} (2013) 076006}
  [\href{https://arxiv.org/abs/1210.5431}{{\ttfamily 1210.5431}}].

\bibitem{Guo:2013sya}
F.-K.~Guo, C.~Hidalgo-Duque, J.~Nieves and M.P.~Valderrama, \emph{{Consequences
  of Heavy Quark Symmetries for Hadronic Molecules}},
  \href{https://doi.org/10.1103/PhysRevD.88.054007}{\emph{Phys. Rev. D}
  {\bfseries 88} (2013) 054007}
  [\href{https://arxiv.org/abs/1303.6608}{{\ttfamily 1303.6608}}].

\bibitem{Liu:2019stu}
M.-Z.~Liu, T.-W.~Wu, M.~Pavon~Valderrama, J.-J.~Xie and L.-S.~Geng,
  \emph{{Heavy-quark spin and flavor symmetry partners of the X(3872)
  revisited: What can we learn from the one boson exchange model?}},
  \href{https://doi.org/10.1103/PhysRevD.99.094018}{\emph{Phys. Rev. D}
  {\bfseries 99} (2019) 094018}
  [\href{https://arxiv.org/abs/1902.03044}{{\ttfamily 1902.03044}}].

\end{thebibliography}\endgroup
